\documentclass[prd,nofootinbib,twocolumn,superscriptaddress,preprintnumbers,balancelastpage,nofootinbib]{revtex4-2}
\usepackage{graphicx}  
\usepackage{dcolumn}   
\usepackage{bm}        
\usepackage{amssymb, amsmath}
\usepackage{slashed}   
\usepackage[colorlinks]{hyperref}
\usepackage{multirow}
\renewcommand{\eqref}[1]{Eq.~(\ref{#1})}

\usepackage{dsfont}

\usepackage[usenames,dvipsnames,svgnames,table]{xcolor}

\def\be{\begin{equation}}
\def\ee{\end{equation}}
\def\bea{\begin{eqnarray}}
\def\eea{\end{eqnarray}}

\newcommand{\mt}{\widetilde{m}_s}
\newcommand{\kmin}{k_{\min}}
\newcommand{\kmax}{k_{\max}}

\newcommand{\fc}{f_\chi}

\newcommand{\oth}{\mathcal{O}^{\mathrm{th}}}
\newcommand{\oexp}{\mathcal{O}^{\mathrm{exp}}}

\def\beq{\begin{equation}}
\def\eeq{\end{equation}}

\def\be{\begin{eqnarray}}
\def\ee{\end{eqnarray}}

\usepackage{graphicx} 

\renewcommand{\vec}{\mathbf}

\newcommand{\bdm}{\begin{displaymath}}
\newcommand{\edm}{\end{displaymath}}

\begin{document}
\widetext

\title{Unveiling dark forces with measurements of the Large Scale Structure of the Universe}
\author{Salvatore Bottaro}
\affiliation{Raymond and Beverly Sackler School of Physics and Astronomy, Tel-Aviv 69978, Israel}
\author{Emanuele Castorina}
\affiliation{Dipartimento di Fisica “Aldo Pontremoli”, Universit\`a degli Studi di Milano and \\INFN, Sezione di Milano,
Via Celoria 16, 20133 Milan, Italy}
\author{Marco Costa}
\affiliation{Scuola Normale Superiore, Piazza dei Cavalieri 7, 56126 Pisa and \\ INFN, Sezione di Pisa, Largo Bruno Pontecorvo 3, 56127 Pisa, Italy and \\
Perimeter Institute for Theoretical Physics, 31 Caroline St N, Waterloo, ON N2L 2Y5, Canada}
\author{Diego Redigolo}
\affiliation{INFN, Sezione di Firenze,
Via Sansone 1, 50019 Sesto Fiorentino, Italy}
\author{Ennio Salvioni}
\affiliation{Dipartimento di Fisica e Astronomia “Galileo Galilei”, Universit\`a di Padova
and \\ INFN, Sezione di Padova,
Via Marzolo 8, 35131 Padua, Italy and \\ Department of Physics and Astronomy, University of Sussex,
Sussex House, Brighton BN1 9RH, UK
}

\begin{abstract}

Cosmology offers opportunities to test Dark Matter independently of its interactions with the Standard Model. We study the imprints of long-range forces acting solely in the dark sector on the distribution of galaxies, the so-called Large Scale Structure (LSS). We derive the strongest constraint on such forces from a combination of Planck and BOSS data. Along the way we consistently develop, for the first time, the Effective Field Theory of LSS in the presence of new dynamics in the dark sector. We forecast that future surveys will improve the current bound by an order of magnitude.
\end{abstract}

\maketitle
\noindent
\section{Introduction}
The overwhelming experimental evidence for Dark Matter (DM) arises purely from its gravitational interactions, which might hide an elaborate dark sector. At the same time, the observed complexity of the interactions of visible matter could suggest that a similar multi-faceted dynamics is at work in the dark sector, with new particle species interacting at different length scales. 

Moreover, the large number of DM production mechanisms~\cite{Preskill:1982cy,Dine:1982ah,Chung:1998zb,Graham:2015rva,Garny:2015sjg} that do not require any non-gravitational portals leaves open the possibility that the dark and visible sectors interact purely through gravity. Under this reasonable (though undesirable) circumstance, precision cosmology and astrophysical probes might be the only tools capable of testing the dynamics of the dark sector.

It is thus of utmost importance to ask how much the nature of the dark sector can be explored through cosmological precision observables \emph{independently} of its non-gravitational portals with the Standard Model. 

Specifically, the question we address in this Letter is whether DM possesses self-interactions affecting cosmological scales. We focus on interactions with range $\lambda_\varphi$ larger than, or comparable to, the size of the Universe today ($\lambda_\varphi > H_0^{-1}$, where $H_0$ is the Hubble constant), with the goal of understanding how strongly present and future cosmological data will test the interaction strength of these long-range dark forces.

We expect the $n$-point correlation functions of galaxies, as measured by ongoing redshift surveys like DESI \cite{DESI:2016fyo} and Euclid \cite{EUCLID:2011zbd}, to be the most suitable observables to test dark forces, since they provide direct information about the behavior of the matter fluctuations. These surveys will collect one order of magnitude more data than the current generation, {\it e.g.}~the BOSS survey~\cite{BOSS:2016wmc}, providing new measurements at an unprecedented precision. An important phenomenological question is thus how much these upcoming datasets will help unveil the hidden dynamics of the dark sector. 

To obtain the theoretical predictions for the galaxy correlators beyond linear Perturbation Theory (PT), we extend the Effective Field Theory of Large Scale Structure (EFTofLSS)~\cite{Baumann:2010tm,Carrasco:2012cv,Perko:2016puo,Foreman:2015lca,Carroll:2013oxa,Carrasco:2013mua,Pajer:2013jj,Senatore:2014vja,Angulo:2014tfa,Baldauf:2014qfa,Porto:2013qua,Vlah:2015sea,Ivanov:2019pdj,Chen:2021wdi}
to include the presence of a dark force acting on DM (see also related studies in the context of modified gravity theories~\cite{Crisostomi:2019vhj,Lewandowski:2019txi,Piga:2022mge}). To our knowledge this is the first consistent perturbative calculation, beyond the linear regime, of the effects of dark sector dynamics on LSS $n$-point functions.

We find that combining the galaxy power spectrum with the Cosmic Microwave Background (CMB) greatly improves the sensitivity to dark forces. Our results are summarized in Fig.~\ref{fig:money}. In particular, our analysis of the two-point correlation function of BOSS data strengthens the constraint from CMB Planck data alone derived in Ref.~\cite{Archidiacono:2022iuu} by a factor of 2. Our forecast for Euclid anticipates an improvement by a factor of 5, leaving open the possibility of finding interesting deviations from $\Lambda$CDM.  

This improvement stems from the distinctive dynamics of dark forces, which leave most of their imprints in the matter power spectrum. From this perspective, our study can be viewed as a novel example of dark sector dynamics that will be uniquely tested by future LSS data, going beyond the present knowledge from Planck measurements of the CMB (see Refs.~\cite{Cyr-Racine:2015ihg,Vogelsberger:2015gpr} for previous approaches). To correctly take into account the interplay between the different observables we find it crucial to employ a particle physics description of the dark sector, which consistently incorporates corrections to both the cosmological background and the fluctuations, thus capturing effects that would be missed by approaches where the two are studied separately.

Finally, our consistent treatment of the EFTofLSS allows us to estimate that further extending the precision of the calculations of the galaxy power spectrum and bispectrum has the potential to improve the reach on dark forces. This observation strongly motivates new theoretical efforts to unleash the full discovery potential of future galaxy surveys.

\section{Setup}\label{sec:setup}

Physically, a long-range dark force can be mediated by an ultralight scalar field with mass $m_\varphi = \lambda_\varphi^{-1} < H_0 \approx 10^{-33}\;\mathrm{eV}$ and trilinear coupling to the DM field $\chi$, $\mathcal{L}_{\text{int}} = -\, \kappa \varphi \chi^2$~\cite{Damour:1990tw,Wetterich:1994bg,Amendola:1999er,Farrar:2003uw}. Introducing the dimensionless field $s = G_s^{1/2} \varphi$, with effective constant $G_s\equiv\kappa^2/m_\chi^4$, the model Lagrangian becomes
\begin{equation}
-2\mathcal{L}\supset (\partial\chi)^2 + m_\chi^2(s)\chi^2 + \left[(\partial s)^2 + m_\varphi^2 s^2\right]/G_s\ ,\label{eq:lag_2}
\end{equation}
where we neglected self-interactions of $s$ and its non-minimal coupling to gravity, which arise at $\mathcal{O}(G_s^{-2})$ and $\mathcal{O}(G_s^{-1} G_N^{-1})$, respectively. The effect of the long-range scalar force can thus be reabsorbed in a space-time dependent mass for DM, $m_\chi(s)=m_\chi (1+2s)^{1/2}$. 

Radiative corrections induced by DM loops make $m_\varphi$ directly sensitive to the DM mass, so that requiring $m_\varphi\lesssim H_0$ bounds the DM mass from above, $m_\chi\lesssim(16\pi^2 M_{\text{Pl}}^2 H_0^2/\beta)^{1/4}$, where we defined
\begin{equation}
\beta \equiv G_s/(4\pi G_N) \label{eq:beta}
\end{equation}
as the strength of the new force normalized to gravity. In the range of $\beta$ of interest for this paper the bound reads $m_\chi\lesssim 10^{-2}\text{ eV}$, favoring a bosonic nature for DM, which we assume to be a scalar for the remainder of this Letter. In light of the current constraint on $\beta$ from the CMB, $\beta \lesssim 0.01$~\cite{Archidiacono:2022iuu}, our analytical results are systematically derived at leading order in the small $\beta$ expansion. 

In this framework the DM can still be described as a collisionless fluid, with geodesics affected by the dark force. Assuming $\chi$ to be pressureless up to nonlinear dynamics, its evolution is governed by continuity and Euler equations with metric perturbations expanded to linear order, whereas the first three moments of the $\chi$ phase space distribution are retained fully nonlinearly~\cite{Baumann:2010tm,Carrasco:2012cv}. 

In turn, the dynamics of the scalar field is dictated by the DM energy density, as discussed in Ref.~\cite{Archidiacono:2022iuu}. At the background level, the Klein-Gordon equation during matter domination has the solution $\bar{s} - \bar{s}_{\text{eq}} = -\, \beta\widetilde{m}_s f_\chi\log ( a/a_{\text{eq}} )$, where $a$ is the scale factor, $\widetilde{m}_s \equiv d\log m_\chi(s)/d s = (1 + 2\bar{s})^{-1}$, $f_\chi\equiv\bar{\rho}_\chi/\bar{\rho}_m$ with $\bar{\rho}_m \equiv \bar{\rho}_\chi + \bar{\rho}_b$, and $ \bar{s}_{\text{eq}}$ is the value of the scalar field at matter-radiation equality, which is found to be practically identical to its initial displacement $\bar{s}_{\rm ini}$ by solving the evolution in radiation domination. 

Throughout cosmic history the light scalar is dominated by kinetic energy, $w_s \simeq +1$, and makes up a subleading fraction of the total energy density. However, the scalar profile modifies the redshift of the DM energy density, accelerating it from $a^{-3}$ to $a^{-(3 + \varepsilon)} a_{\rm eq}^\varepsilon$, which in turn modifies cosmological distances according to 
\begin{equation}
\frac{H(a)}{H_{\rm CDM}(a)}  \simeq 1 - \frac{1}{2}\varepsilon f_\chi \bigg( \hspace{-0.75mm}\log \frac{a}{a_{\rm eq}} - \frac{1}{3}\bigg)\, , \quad\varepsilon \equiv \beta  \widetilde{m}_s^2 f_\chi\ , \label{eq:distances}
\end{equation}
as discussed in Appendix~\ref{sec:eulerian}. We assume that the scalar has negligible initial displacement, $\bar{s}_{\rm ini} \simeq 0$, yielding $\widetilde{m}_s \simeq 1$ and $\varepsilon \simeq \beta f_\chi$ at leading order in $\beta$. We further assume that all of DM couples to the long-range force, though we comment in Sec.~\ref{sec:discussion} on scenarios where only a fraction of the DM energy density is interacting.

\section{Signatures in cosmological correlators}\label{sec:correlators}
The presence of a scalar long-range force modifies the sub-horizon dynamics in two ways: i) it enhances the growth of the total matter fluctuations; ii) it sources relative density and velocity fluctuations between DM and baryons. In this section we demonstrate why the first effect dominates in scenarios where the dark force interacts with the totality of DM (\emph{i.e.}~$f_\chi\simeq1$). We leave a discussion of scenarios with $f_\chi\ll1$ for a companion study~\cite{MultitracerPaper}.

The observables measured in redshift surveys are correlation functions of the galaxy number density perturbation $\delta_g \equiv n_g/\bar{n}_g - 1$, which, at long wavelengths, can be expanded in terms of the matter fields as~\cite{Desjacques:2016bnm,Schmidt:2016coo,Chen:2019cfu}
\begin{equation}\label{eq:bias_first_text}
\delta_g = b_1 \delta_m + b_r \delta_r + b_\theta \theta_r + \mathrm{nonlinear}\;\mathrm{terms}\,,
\end{equation}
where $\theta \equiv \nabla_i v^i$. In the above we defined the density contrasts $\delta_x \equiv \rho_x / \bar{\rho}_x - 1$ for $x = \chi, b$, the total and relative matter perturbations 
\begin{equation}
\delta_m = f_\chi \delta_\chi + (1 - f_\chi) \delta_b\,,\qquad \delta_r = \delta_\chi - \delta_b\, ,
\end{equation}
and similarly for the velocity fields $\vec{v}_m$ and $\vec{v}_r$.

\subsection{Enhanced growth}\label{sec:growth}
In the sub-horizon limit $k \gg a H$ and deep in matter domination, the evolution equations for the total matter perturbations maintain the same structure of the CDM ones. This follows from the absence of new scales in scenarios with long-range dark forces. At linear level we find $\delta_m^{(1)} (\vec{k},a) = D_{1m} \delta_0(\vec{k})$, with a linear growth factor
\begin{equation}\label{eq:D1m}
D_{1m} \simeq \left[1 + \varepsilon f_\chi \left( \log \frac{a}{a_{\rm eq}} - \frac{7}{3}\right)\right]D_{1m}^{\rm CDM}\,, 
\end{equation}
where $D_{1m}^{\rm CDM} = a$ and $\delta_0(\vec{k})$ is related to the value of the fluctuation at matter-radiation equality. The overall growth gets enhanced compared to the CDM case at $\mathcal{O}(\beta)$, with a large logarithm boosting this effect even further. This logarithm has the same origin of the one that appeared in \eqref{eq:distances} and it is connected to the dark force being long-range throughout the evolution of the Universe.

At the nonlinear level we find a remarkably simple structure for the solutions at order $n\geq 2$,
\begin{equation}\label{eq:deltam_nonlinear}
\delta_m^{(n)} (\vec{k}, a) = (D_{1m})^n \hspace{-1mm}\int_{\vec{k}} \mathrm{d}k_{1 \ldots n} [F_n + \varepsilon f_\chi \Delta F_n] (\vec{k}_1, \ldots, \vec{k}_n)  \,,
\end{equation}
where $F_n$ is a standard CDM kernel~\cite{Bernardeau:2001qr} and $\int_{\vec{k}} \mathrm{d}k_{1 \ldots n} \hspace{-0.2mm}\equiv\hspace{-0.2mm} \int \frac{\mathrm{d}^3 k_1 \ldots \mathrm{d}^3 k_n} {(2\pi)^{3(n-1)}} \,\delta^{(3)}\big(\vec{k} - \sum_{i\, =\, 1}^n \vec{k}_i \big)\delta_0 (\vec{k}_1) \ldots \delta_0 (\vec{k}_n)$.
Modulo non-log-enhanced subleading terms, the time dependence is fully encapsulated by the linear growth factor, as in $\Lambda$CDM. Furthermore, the new nonlinearities induced by the dark force, $\Delta F_n$, do not introduce any additional spatial structure and do not change the infrared (IR) behavior of the solution (explicit expressions can be found in Appendices~\ref{sec:eulerian} and~\ref{sec:lagrangian}). 

The form of~\eqref{eq:deltam_nonlinear} greatly simplifies the evaluation of the nonlinear dynamics, allowing us to model the effects of new physics on cosmological correlators at $\mathcal{O}(\varepsilon f_\chi\log a/a_{\text{eq}})$. In this approximation, $\delta_0(\vec{k})$ is related to the inflationary perturbation through the $\Lambda$CDM transfer function~\cite{Dodelson:2003ft}.

With this model at hand, we can easily derive a one-dimensional Fisher forecast on the dark force coupling, which estimates the idealized Euclid reach (see Appendices~\ref{sec:methods} and \ref{sec:full_results} for details). At $95\%$~c.l.~from the real space power spectrum we find $(\varepsilon f_\chi)_{P_g} < 1.8\times 10^{-4}$ for $k_{\text{max}} = 0.27\, h\, \text{Mpc}^{-1}$. Realistic constraints in redshift space from BOSS and forecasts accounting for degeneracies will be presented in Sec.~\ref{sec:results}.

\subsection{Relative densities and velocities}\label{sec:EPtests}

The relative perturbations are proportional to $\varepsilon \delta_m$ at linear order, and their growth is not enhanced by large logs at $\mathcal{O}(\beta)$: $\delta_r^{(1)} = -\, \theta_r^{(1)}/(a H) = 5 \varepsilon \delta_m^{(1)}/3$. This remains true nonlinearly, where
\begin{equation}
\delta_r^{(n)} (\vec{k}, a) = \varepsilon\, \big(D_{1m}^{\rm CDM}\big)^n \int_{\vec{k}} \mathrm{d}k_{1 \ldots n} F_{nr} (\vec{k}_1, \ldots, \vec{k}_n)  \,.
\end{equation}
Therefore, to account for the leading log-enhanced effects when $f_\chi \simeq 1$ it is sufficient to retain only the total matter perturbations in the bias expansion of~\eqref{eq:bias_first_text}. The relative error associated to this approximation is of order $10\%/f_\chi$.

Nevertheless, the presence of large scale relative perturbations effectively breaks the Equivalence Principle (EP) and can be tested by looking at the breakdown of the consistency relations which constrain the IR structure of cosmological correlation functions in $\Lambda$CDM~\cite{Kehagias:2013yd,Peloso:2013zw,Peloso:2013spa,Creminelli:2013cga,Creminelli:2013mca,Creminelli:2013poa,Creminelli:2013nua,Kehagias:2013rpa,Valageas:2013cma,Horn:2014rta,Horn:2015dra,Inomata:2023faq}.\footnote{The assumption of adiabatic initial conditions could be broken, depending on the dynamics of the dark force field $s$ during inflation. We point the reader to Ref.~\cite{Archidiacono:2022iuu} for a model-independent discussion.}

At the level of the galaxy power spectrum $P_g$, the EP ensures that bulk flows do not contribute to shifts in the position of the BAO scale~\cite{Jain:1995kx,Scoccimarro:1995if,2012PhRvD..85j3523S}, which will be measured down to few permille by future surveys. New operators in the galaxy bias expansion \cite{Dalal:2010yt,Blazek:2015ula,Chen:2019cfu,Givans:2020sez} can in general cause a shift, but their effect is suppressed in our model because the relative matter fluctuations follow the total matter ones at linear level: $\delta_r^{(1)} \propto \beta \delta_m^{(1)}$ and $\mathbf{v}_r^{(1)} \propto \beta \mathbf{v}_m^{(1)}$.

At higher order in the EFT, and in the presence of an EP violation, one would naively expect $\mathcal{O}(\beta)$ shifts of the BAO position. However, as shown in Appendix~\ref{sec:consistency_rel}, at the power spectrum level relative motions only enter as $\mathbf{v}_r^2\sim \mathcal{O}(\beta^2)$ and are therefore negligible, given the CMB constraint on $\beta\lesssim0.01$.\footnote{We neglect relative velocity fluctuations sourced by the tight coupling between baryons and photons before recombination, which decay as $1/a$.} We explicitly checked the cancellation of bulk flows at one loop and $\mathcal{O}(\beta)$ in the matter and galaxy power spectra.\footnote{Appendix~\ref{sec:consistency_rel} provides two arguments
for why this cancellation must take place.} The same logic applies to higher-order bias parameters in the relative velocities and densities. 

Beyond the two-point function, EP violation can reveal itself in the form of a $1/p$ pole in the squeezed limit of the bispectrum $B_g$, where $p$ is the long wavelength mode~\cite{Kehagias:2013yd,Peloso:2013zw,Peloso:2013spa,Creminelli:2013cga,Creminelli:2013mca,Creminelli:2013poa,Creminelli:2013nua,Crisostomi:2019vhj,Lewandowski:2019txi}. For fluctuations of two different tracers, $\delta_g^A$ and $\delta_g^B$, and using the bias expansion in Eq.~(\ref{eq:bias_first_text}), the squeezed limit of the bispectrum  reads
\begin{align}
&\frac{B^{AAB}_g (\vec{p}, \vec{p}_1, \vec{p}_2 )}{P_{m, L}^{\rm CDM} (p) P_{m, L}^{\rm CDM} (p_1)}\bigg|_{\vec{p}\, \to \, 0} \simeq \varepsilon\hspace{0.3mm} \frac{\vec{p}\cdot \vec{p}_1}{p^2} \frac{7b_1^A}{6} \Delta b^{AB}\,,  \label{eq:pole_B_real}
\end{align}
where $\Delta b^{AB}\equiv b_1^A \overline{b}_r^B - b_1^B \overline{b}_r^A $ and $\overline{b}_r \equiv b_r - 4 \mathcal{H} b_\theta/7$ up to $\mathrm{non}$-$\mathrm{linear}\;\mathrm{terms}$ discussed in Appendix~\ref{sec:eulerian}. For a single tracer, $A=B$, the pole vanishes due to the symmetry under exchange of the two short modes. The above expression generalizes the results of Refs.~\cite{Creminelli:2013nua,Peloso:2013spa}, by consistently including the modifications to the cosmological background and all the relevant bias operators.

Analytical and numerical estimates of relative bias parameters~\cite{Schmidt:2016coo,Barreira:2019qdl} indicate that, in a $\Lambda$CDM model, $b_r\,, \mathcal{H}b_\theta \sim \mathcal{O}(1)$. A one-dimensional Fisher forecast gives at 95\% c.l.~$(\varepsilon f_\chi^{1/2})_{B_g^{AAB}} < 1.0\times 10^{-2}\, \big(b_1/(\overline{b}_r^A - \overline{b}_r^B)\big)^{1/2}$ for $b_1^A=b_1^B=b_1$ and $k_{\text{max}}=0.11\, h\, \text{Mpc}^{-1}$. Even an optimistic reach for maximally different tracers is much weaker than the sensitivity of the power spectrum, although with a different scaling with $f_\chi$ (see Appendix~\ref{sec:methods} for a detailed derivation).

We therefore conclude that, for $f_\chi \simeq 1$, the impact of dark forces on the galaxy power spectrum and bispectrum is dominated by the log-enhanced growth of structure discussed in Sec.~\ref{sec:growth}. We focus on this effect from now on. 

\section{Constraints and Discussion}~\label{sec:results}

Given the results discussed in the previous sections, we are now in the position to search for long-range interactions in the dark sector with galaxy survey data. In this work we make use of the BOSS data release 12~\cite{BOSS:2016wmc}. This contains approximately one million galaxies, divided in two redshift bins, with $z_{\rm eff} = 0.32$ and $0.57$, and in two patches, above and below the galactic hemisphere. We include measurements of the position of the reconstructed BAO from BOSS~\cite{BOSS:2016wmc} and other surveys~\cite{Kazin:2014qga,Beutler:2011hx,Ross:2014qpa}, and Planck measurements of the CMB power spectra, including lensing~\cite{Planck:2018vyg}. The theoretical model for the power spectrum is evaluated with the \texttt{CLASS}~\cite{Blas:2011rf,Lesgourgues:2011re} and \texttt{PyBird}~\cite{DAmico:2020kxu,PyBird} codes and compared to the data up to $k_{\rm max} = 0.20~(0.23)\, h\,\mathrm{Mpc}^{-1}$ for the lower~(higher) redshift bin. We sample the posterior of the cosmological and nuisance parameters with \texttt{MontePython} \cite{Audren:2012wb,Brinckmann:2018cvx}. The final constraint on $\beta$ is always marginalized over the $6$ standard cosmological parameters and $8$ nuisance parameters in the galaxy power spectrum analysis, per redshift bin and galactic cap, to account for nonlinear galaxy bias, shot-noise, and the counterterms of the EFT.

\begin{figure}
\includegraphics[width=0.475\textwidth]{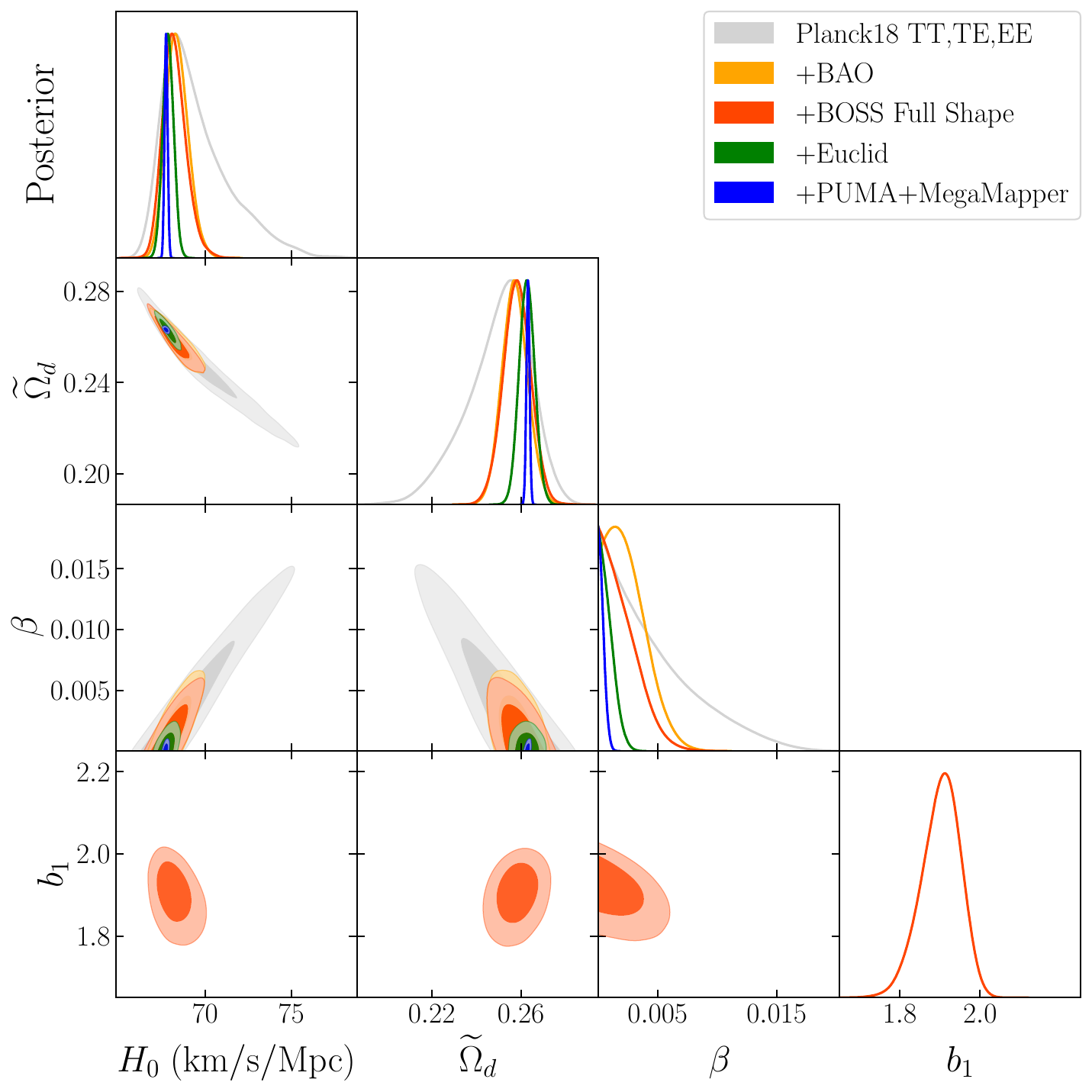}
\caption{Present constraints and forecasted reach on the dark force parameter space, obtained from the analysis of the redshift space galaxy power spectrum using the EFTofLSS. $\beta$ is the strength of the dark force normalized to gravity, as defined in~\eqref{eq:beta}. $\widetilde{\Omega}_d$ serves as a proxy for $\Omega_\chi^0$, see the main text for details. $b_1$ is the linear bias parameter for BOSS galaxies with $z_{\rm eff} = 0.57$ and in the north galactic cap. Gray~(orange) contours show the constraints from Planck~(Planck+BAO), previously obtained in Ref.~\cite{Archidiacono:2022iuu}. Red contours show the constraints after including the analysis of the full shape of the BOSS power spectrum~\cite{BOSS:2016wmc}. Green contours correspond to our forecast for the ongoing Euclid survey, while blue contours include the proposed PUMA and MegaMapper surveys.\label{fig:money}}
\end{figure}

A summary of the bounds is shown in Fig.~\ref{fig:money}, where we plot the 4-dimensional parameter space of $\beta$, the Hubble constant $H_0$, the parameter $\widetilde{\Omega}_d$, serving as a proxy for $\Omega_\chi^0$ and defined as the fractional energy density the interacting DM $\chi$ would have today had it evolved like $a^{-3}$~\cite{Archidiacono:2022iuu}, and the linear bias of the high-redshift bin for the north galactic cap. At 95\% c.l. we obtain $\beta< 0.012$ from Planck data (gray contours), which improves to $\beta< 0.0048$ including BOSS full shape information on top of all the present BAO data (red contours). 

The degeneracy between $H_0$ and $\beta$ moves along the direction that keeps the angular size of the sound horizon at recombination constant: increasing $\beta$ reduces the amount of matter at late times and can be compensated by a larger $H_0$ (or, equivalently, smaller $\widetilde{\Omega}_d$). The addition of the full shape of the BOSS galaxy power spectrum does not significantly improve over the Planck+BAO constraint presented in Ref.~\cite{Archidiacono:2022iuu}. This was expected, since the dominant effect of the new force is an approximately constant upward shift of the amplitude of the matter power spectrum, which can be compensated by decreasing the linear bias $b_1$, defined in~\eqref{eq:bias_first_text}. Given the bound from Planck+BAO on $\beta$, and the statistical errors of the BOSS data, the uncertainty on the linear bias is too large to provide significant additional constraining power. In other words, the modification of the background evolution induced by the dark force, as probed by CMB and BAO data, trumps the enhanced growth of structure at late times, which at the power spectrum level is highly contaminated by the nuisance parameters (see Appendix~\ref{sec:full_results} for the complete results). 

We then use our theoretical model to forecast the reach of future spectroscopic surveys and $21\,\mathrm{cm}$ instruments, such as the ongoing DESI and Euclid, and the proposed MegaMapper~\cite{Schlegel:2019eqc} and PUMA~\cite{PUMA:2019jwd}. Our forecasts are based on the 1-loop modeling of the redshift space galaxy power spectrum presented in Ref.~\cite{Chen:2020zjt} and implemented in the Fisher forecasting code \texttt{FishLSS}~\cite{Sailer:2021yzm,FishLSS}.\footnote{We proceed as follows. We invert the covariance matrix of the Planck+BOSS Markov Chain Monte Carlo (MCMC) chains to obtain the associated Fisher matrix and we add it to the Fisher matrix of the forecasts, computed with \texttt{FishLSS}. Next, we generate a Gaussian sample centered at the best fit value of the MCMC (putting $\beta = 0$) with covariance equal to the inverse of the combined Fisher matrix. Finally, we drop all points with negative $\beta$ and derive the confidence intervals.} The green lines in Fig.~\ref{fig:money} show the improvement in the bound on $\beta$ brought by adding the Euclid spectroscopic sample to the Planck and BOSS datasets, while the blue line corresponds to the further inclusion of MegaMapper and PUMA. We find that future surveys, thanks to their large volume, could dramatically improve current bounds and reach $\beta < 0.002\, (0.0008)$ at 95\% c.l.~for Euclid~(PUMA+MegaMapper), even after marginalizing over all the nuisance parameters of the nonlinear modeling. 

As we detailed above, the strong degeneracy between galaxy bias and $\beta$ limits the ultimate reach of a power spectrum analysis. It is thus interesting to see to what extent higher order correlation functions such as the bispectrum, which is known to help break degeneracies between cosmological and nuisance parameters~\cite{Sefusatti:2006pa,Rizzo:2022lmh,Ivanov:2023qzb,DAmico:2022osl}, can tame this effect and provide stronger bounds on DM long-range interactions. 

As a proof of concept we can forecast the constraining power of a joint analysis for the Euclid survey \cite{Euclid:2019clj}, combining the real space power spectrum at one loop with the real space bispectrum at tree level. For the latter we include all triangular configurations up to $k_{\rm max} = 0.11\,h\,\mathrm{Mpc}^{-1}$, independently of redshift. We find that the limited $k$ range imposed by our tree-level calculation prevents the bispectrum from helping significantly in breaking parameter degeneracies, once priors derived from current data are imposed. From a single-parameter Fisher forecast we find at $95\%$~c.l.~$(\varepsilon f_\chi)_{B_g} < 8.7\,\times 10^{-4} \left( 0.11\,h\,\mathrm{Mpc}^{-1}/k_{\rm max}\right)^{2.2}$ for the bispectrum alone (see Appendices~\ref{sec:methods} and \ref{sec:full_results} for further details). The dependence of the bispectrum reach on $k_{\text{max}}$ shows that extrapolating to $\approx 0.2\,h\,\mathrm{Mpc}^{-1}$ could yield a comparable sensitivity to the power spectrum one presented in Sec.~\ref{sec:growth}. We view this as a strong indication that unleashing the power of higher order correlators to constrain dark forces requires a 1-loop modeling of the bispectrum, and possibly the inclusion of the trispectrum~\cite{Sefusatti:2004xz,Bertolini:2016bmt,Crisostomi:2019vhj,Lewandowski:2019txi}.

As the uncertainty on $\beta$ decreases, higher order terms in the perturbative series for $\Lambda$CDM, \emph{e.g.}~2-loop corrections to the power spectrum, can become comparable in size to new physics effects. Using estimates based on available results~\cite{Nishimichi:2020tvu} we find that constraints on $\beta$ down to $\mathcal{O}(10^{-3})$ are still insensitive to the theory error. This issue is howeever more pressing for the bispectrum and could even affect future analyses of the power spectrum, if $\beta \sim \mathcal{O}(10^{-4})$ sensitivity will be achieved.

\section{Outlook}~\label{sec:discussion}
In this Letter we derived the first constraints on long-range forces acting on DM from the LSS of the Universe. We showed that including the galaxy power spectrum as currently measured by BOSS improves the constraint on the dark force strength by a factor of 2 with respect to the CMB alone, reaching $\beta < 0.0048$ at $95\%$ c.l.. This is the strongest bound to date on long-range interactions in the dark sector from cosmological data (see Refs.~\cite{Kesden:2006vz,Desmond:2020gzn} for constraints from galactic dynamics). A further tightening is forecasted with upcoming measurements from larger surveys, for example \mbox{$\beta < 0.002$} for Euclid.

This work can be expanded in several directions. A pressing one is to extend the perturbative modeling of higher-point correlators in the presence of dark forces. In particular, we found that the galaxy bispectrum may reach the same level of sensitivity of the power spectrum. Beyond galaxy $n$-point functions, probing directly the galaxy velocity field,~\emph{e.g.}~with peculiar velocity surveys~\cite{Gordon:2007zw,Howlett:2022len,Qin:2019axr}, could enable auxiliary constraints and help isolate the non-log-enhanced EP violations.

In addition, other regions of the parameter space of dark forces can be probed using the results presented here. First, if only a small fraction of DM is self-interacting ({\it i.e.}~$f_\chi \ll 1$) the contributions to galaxy correlators of relative perturbations, proportional to $\beta f_\chi$, can overcome the log-enhanced corrections, which scale as $\beta f_\chi^2 \log a/a_{\text{eq}}$. The interplay of these effects in testing the $f_\chi \ll 1$ scenario will be discussed in a forthcoming publication~\cite{MultitracerPaper}. Second, one is naturally led to consider heavier mediators, $m_\varphi \gtrsim H_0$, corresponding to a shorter range for the dark force. In this regime, the evolution of the cosmological background is modified, as the mediator itself is expected to constitute a fraction of DM today. Addressing these questions will allow us to form a complete picture of what LSS can teach us about DM self-interactions. 

\acknowledgments
We thank Maria Archidiacono, Diego Blas, Guido D'Amico, Matt Lewandowski, Noah Sailer, Marko Simonović, and Filippo Vernizzi for helpful discussions. We are grateful to Pierre Zhang for guidance and support in the use of the \texttt{PyBird} code. We further thank Tomer Volansky and Martin White for comments on the draft. We thank Cyril Creque-Sarbinowski, Olivier Simon and Zachary Weiner for
pointing out that in the previous v2 some coefficients in the analytical discussion of subhorizon solutions were incorrect. SB is supported by the Israel Academy of Sciences and Humanities \& Council for Higher Education Excellence Fellowship Program for International Postdoctoral Researchers. MC was supported in part by Perimeter Institute for Theoretical Physics. Research at Perimeter Institute is supported by the Government of Canada through the Department of Innovation, Science and Economic Development Canada and by the Province of Ontario through the Ministry of Research, Innovation and Science. E.S. acknowledges partial support from the EU’s Horizon 2020 programme under the MSCA grant agreement 860881-HIDDeN, and was supported in part by the Science and Technology Facilities Council under the Ernest Rutherford Fellowship ST/X003612/1.

\bibliographystyle{JHEP}
\bibliography{5thLSS.bib}

\clearpage
\newpage
\appendix
\onecolumngrid
\section{Effective field theory of long-range dark forces}\label{sec:eulerian}

The DM dynamics in the presence of the dark force can be treated as that of a collisionless fluid, where the effect of the dark force modifies the DM geodesics. Assuming $\chi$ to be pressureless up to its nonlinear dynamics and expanding to the linear order in the metric perturbations $\Psi, \Phi$, we can write the continuity and Euler equations in Newtonian gauge as
\begin{equation}
\begin{split}
   \rho_\chi' + \big(3(\mathcal{H} + \Phi')- \widetilde{m}_ss'\big) \rho_\chi + \nabla_i( \rho_\chi v_\chi^i)  &= 0\, , \\
 (\rho_\chi v_\chi^i)^{\prime} + 4 \mathcal{H} \rho_\chi v_\chi^i +\rho_\chi\nabla^i (\Psi + \widetilde{m}_s s) + \nabla_j \Sigma^{ij}_\chi &= 0\, ,
 \end{split}
 \label{eq:fulleq}
\end{equation}
where primes denote derivatives with respect to conformal time $\tau$, $\mathcal{H}=a'/a$ is the conformal Hubble parameter, and we defined $\widetilde{m}_s \equiv d\log m_\chi(s)/d s$. To describe the matter fluctuations at the level of precision required here it is sufficient to know the DM energy density $\rho_\chi$, the DM fluid velocity $v^i_\chi$, and the second moment of the DM phase space distribution $\Sigma^{ij}_\chi$~\cite{Carrasco:2012cv}. The higher moments of the DM phase space distribution produce subleading corrections in the $p/E$ expansion and can be neglected.  

At the background level, the DM density sources a scalar field profile through the Klein-Gordon equation
\begin{equation}
\bar{s}''+2\mathcal{H} \bar{s}' + G_s a^2\widetilde{m}_s\bar{\rho}_\chi=0\ ,\label{eq:sbkd}
\end{equation}
where the scalar field mass $m_\varphi \lesssim H_0$ can be safely ignored throughout the cosmological evolution. The cubic interaction $\mathcal{L}_{\text{int}} = -\, \kappa \varphi \chi^2$ gives $\widetilde{m}_s = (1 + 2\bar{s})^{-1}$, which ensures that the scalar profile evolves following the DM source term \emph{independently} of its initial displacement, as long as $\bar{s}_{\text{ini}}\lesssim 1$.\footnote{A polynomial interaction of the form $\mathcal{L}_{\text{int}} = -\,\kappa_n \varphi^n \chi^2$ leads to $m_\chi(s)=m_\chi (1 + 2 s^n)^{1/2}$ and $\widetilde{m}_s = n \bar{s\,}^{n-1}/(1 + 2 \bar{s}^{\,n})$. For $n>1$ the coupling strength of the dark force is suppressed by the initial scalar field displacement, which in this work is assumed to satisfy $\bar{s}_{\text{ini}} \ll 1$. An upper bound on $\bar{s}_{\text{ini}}$ can be derived by requiring the potential energy to be negligible throughout the scalar field evolution: $\bar{s}_{\text{ini}} \ll \sqrt{3\beta}\, H_0/m_\varphi$. For the scalar masses considered here, this requirement rules out large initial field displacements. Since any non-polynomial interaction can be expanded in a polynomial basis, we believe that our discussion of the cubic interaction with $n = 1$ should capture the generic consequences of the presence of a long-range dark force. By contrast, for $\bar{s}_{\text{ini}}\gg 1$ the dark force effectively decouples as $\widetilde{m}_s\sim 1/\bar{s}_{\text{ini}}$, independently of the microscopic interaction.} In turn, the presence of the dark force modifies the redshift of the DM energy density. Writing the continuity equation  
\begin{equation}
\bar{\rho}_\chi'+\left(3\mathcal{H}-\widetilde{m}_s \bar{s}'\right)\rho_\chi=0\ ,\label{eq:chibkd}
\end{equation}
one can see that the background scalar field variation shifts the Hubble friction compared to the $\Lambda$CDM one. For convenience we define the fraction of $\chi$ as $f_\chi\equiv\bar{\rho}_\chi/\bar{\rho}_m$, with $\bar{\rho}_m \equiv \bar{\rho}_\chi + \bar{\rho}_b$. In matter domination and at $\mathcal{O}(\beta)$ we can take $\mathcal{H} = 2/\tau$ and $a = \mathcal{C} \tau^2$ with $\mathcal{C} \equiv \Omega^0_m H_0^2/4$ in~\eqref{eq:sbkd} to solve for the scalar field profile, obtaining $\bar{s} - \bar{s}_{\text{eq}} = -\,2\beta\widetilde{m}_s f_\chi\log(\tau/\tau_{\text{eq}})$. Via~\eqref{eq:chibkd}, the decreasing value of $\bar{s}$ makes the interacting DM $\chi$ redshift \emph{faster} than cold DM by a factor of $2\varepsilon \log(\tau/\tau_{\text{eq}})$, where we defined $\varepsilon \equiv \beta f_\chi \widetilde{m}_s^2$ as in~\eqref{eq:distances}. As a consequence, the Hubble parameter in matter domination also receives a log-enhanced correction, $H(\tau) / H_{\rm CDM} (\tau) \simeq 1 + 2 \varepsilon f_\chi [ \log (\tau / \tau_{\rm eq}) - 5/3 ]$, which modifies cosmological distances. The conformal Hubble parameter has the expression $\mathcal{H} =  (1 -  \varepsilon f_\chi) 2/\tau$.

Defining the density contrasts $\delta_x = \rho_x / \bar{\rho}_x - 1$ for $x = \chi, b$, it is convenient to switch variables to the total and relative matter perturbations
\begin{equation} \label{eq:deltam_deltar}
\delta_m = f_\chi \delta_\chi + (1 - f_\chi) \delta_b\,,\qquad \delta_r = \delta_\chi - \delta_b\,,
\end{equation}
and similarly for $\vec{v}_m$ and $\vec{v}_r$. Neglecting the different initial conditions between DM and baryons generated by the baryon-photon coupling, $\chi$ and $b$ have identical evolution during matter domination in the $\beta \to 0$ limit. This implies that all relative fluctuations $\delta_r, \,\vec{v}_r$ are $\mathcal{O}(\beta)$. In light of the strong constraints on $\beta$ from linear theory derived in Ref.~\cite{Archidiacono:2022iuu}, also at the level of the perturbations we systematically expand for small $\beta$, always dropping $\mathcal{O}(\beta^2)$ terms. 

The evolution equations for the total matter perturbations and the relative perturbations in the sub-horizon limit $k \gg \mathcal{H}$ and deep in matter domination can be derived from \eqref{eq:fulleq} and read 
\begin{equation}
\begin{split}
&\delta_m^\prime + \theta_m =\, - \nabla_i (\delta_m v_m^i)\,, \\
&\theta_m^\prime + \mathcal{H}(1 - f_\chi \varepsilon) \theta_m + \frac{3}{2}\Omega_m \mathcal{H}^2 \delta_m ( 1 + f_\chi \varepsilon) = -\nabla_i (v_m^j \nabla_j v_m^i)\, ,\label{eq:tot_pert}\\
&\delta_r^\prime + \theta_r =\, - \nabla_i (\delta_m v_r^i + \delta_r v_m^i )\, , \\
&\theta_r^\prime + \mathcal{H} \theta_r - \varepsilon \mathcal{H} \Big(  \theta_m -  \frac{3}{2} \Omega_m \mathcal{H}   \delta_m \Big) =\,  -\nabla_i (v_m^j \nabla_j v_r^i)- \nabla_i (v_r^j \nabla_j v_m^i)\ ,
\end{split}
\end{equation}
where $\theta_{m,r} \equiv \nabla_i v^i_{m, r}$. These are the Standard Perturbation Theory (SPT) equations, whose solutions can be used to model the matter power spectrum and the bispectrum on linear and nonlinear scales. Adiabatic initial conditions for the perturbations are assumed~\cite{Archidiacono:2022iuu}.

At first order in the fields, the bias expansion of the galaxy number density perturbation $\delta_g \equiv n_g/\bar{n}_g - 1$ reads~\cite{Schmidt:2016coo,Chen:2019cfu}
\begin{equation}\label{eq:bias_first}
\delta_g = b_1 \delta_m + b_r \delta_r + b_\theta \theta_r\,.
\end{equation}
The linear solutions of Eqs.~(\ref{eq:tot_pert}) are given by
\begin{equation} \label{eq:linear_SPT}
\delta_{m,r}^{(1)} (\vec{k}, a) = D_{1m,1r}\,\delta_0 (\vec{k})\,, 
\end{equation}
in terms of the growth factors, which for $a\gg a_{\rm eq}$ have the expressions\footnote{For the total matter perturbation we impose the initial conditions $\delta_m^{(1)}(\vec{k}, \tau_{\rm eq}) = \delta_m^{(1),\mathrm{CDM}}(\vec{k}, \tau_{\rm eq})$ and $\delta_m^{(1) \prime}(\vec{k}, \tau_{\rm eq}) = \delta_m^{(1),\mathrm{CDM}\, \prime}(\vec{k}, \tau_{\rm eq})$, where $\delta_m^{(1),\mathrm{CDM}}(\vec{k}, \tau) \equiv D_{1m}^{\rm CDM}(\tau) \delta_0(\vec{k})$. For the relative matter perturbation we impose $\delta_r^{(1)}(\vec{k}, \tau_{\rm eq}) = \delta_r^{(1)\prime}(\vec{k}, \tau_{\rm eq}) = 0$.}
\begin{equation}
D_{1m} \simeq \left[1 + \varepsilon f_\chi \left( \log \frac{a}{a_{\rm eq}} - \frac{7}{3}\right)\right]D_{1m}^{\rm CDM}(a) \,, \qquad
D_{1r} \simeq\, \frac{5}{3}\,\varepsilon D_{1m}^{\rm CDM} (a) \,,
\end{equation}
where $D_{1m}^{\rm CDM}(a) = a$ is the standard growth factor in Einstein-de Sitter. The linear solutions for the velocity divergences read $\theta_{m,r}^{(1)} = - f_{m, r} \mathcal{H}\, \delta_{m,r}^{(1)}\,$, where for the growth rates $f_{m,r} = d\log D_{1m,1r}/d\log a$ we find $f_m = 1 + \varepsilon f_\chi$ and $f_r = 1$, again working up to $\mathcal{O}(\beta)$.

For completeness we also provide the linear growth factors as functions of conformal time,
\begin{equation}
D_{1m} \simeq \, D_{1m}^{\rm CDM}(\tau) \,, \qquad D_{1r} \simeq \, \frac{5}{3}\,\varepsilon D_{1m}^{\rm CDM}(\tau)\,,
\end{equation}
where $D_{1m}^{\rm CDM}(\tau) = \mathcal{C}\tau^2$. The relation between the scale factor and conformal time is for $\tau \gg \tau_{\rm eq}$
\begin{equation}
a(\tau) \simeq \mathcal{C}\tau^2 \left( 1 - 2 \varepsilon f_\chi \log \frac{\tau}{\tau_{\rm eq}} + \frac{7}{3}\varepsilon f_\chi \right)\,.
\end{equation}
We thus find from~\eqref{eq:bias_first} that $\delta_g^{(1)}(\vec{k}, a) = D_{1m} \big( b_1 + \varepsilon\hspace{0.4mm} 5 \widehat{b}_r /3  \big) \delta_0 (\vec{k}),$ where $\widehat{b}_r \equiv b_r - b_\theta \mathcal{H}$. The galaxy power spectrum at tree level reads, retaining only the leading log-enhanced terms,
\begin{equation}
P_{g, L}(k) \simeq  b_1^2 \left( 1 + 2 \varepsilon f_\chi \log \frac{a}{a_{\rm eq}} \right) P_{m,L}^{\rm CDM} (k)\,,  
\end{equation}
with $P_{m,L}^{\rm CDM} (k) = \big(D_{1m}^{\rm CDM}(a) \big)^2 P(k)$. The power spectrum at early reference time is defined by $\langle \delta_0 (\vec{k}) \delta_0 (\vec{k}^\prime) \rangle = (2\pi)^3 \delta^{(3)} (\vec{k} + \vec{k}^\prime) P(k)$. This shows that the leading effect of the dark force on the power spectrum at low redshift is a scale-independent enhancement proportional to $\varepsilon f_\chi \log\, [ (1 + z_{\rm eq})/(1 + z) ] \approx 8\hspace{0.3mm}\varepsilon f_\chi$~\cite{Archidiacono:2022iuu}.

Up to second order in the fields and $\mathcal{O}(\beta)$, the bias expansion is extended from~\eqref{eq:bias_first} to~\cite{Schmidt:2016coo,Chen:2019cfu} \vspace{-1.5mm}
\begin{equation}
\begin{split}\label{eq:bias_second}    
&\;\delta_g = b_1 \delta_m + b_r \delta_r + b_\theta \theta_r + \frac{b_2}{2}\delta_m^2 + b_{K^2} K_{ij} K^{ij} + b_{mr} \delta_m \delta_r + b_{\delta\theta} \delta_m \theta_r + b_{\nabla \delta} \nabla_i \delta_m  v_r^i + b_K K_{ij} \nabla^i v_r^j + b_{Kr} K_{ij} \frac{\nabla^i \nabla^j}{\nabla^2} \delta_r
\end{split}
\end{equation}

\vspace{-4mm}\noindent where $K_{ij} \equiv \big( \nabla_i \nabla_j/\nabla^2 - \delta_{ij}/3\big)\delta_m\,$ is the traceless tidal tensor. The above expansion should also contain terms like $b_{v_r^2} v_r^2$ which, however, are of $\mathcal{O}(\beta^2)$ in our counting and therefore negligible. We notice that the last operator in the above expansion is missing in previous work on the bias expansion in presence of relative density and velocity perturbations \cite{Schmidt:2016coo,Chen:2019cfu}. 
The second order solutions of Eqs.~(\ref{eq:tot_pert}) are given by
\begin{equation}
\label{eq:second_order}
    \delta_m^{(2)} (\vec{k},a) =  (D_{1m})^2 \int_{\vec{k}} \mathrm{d}k_{12}\widetilde{F}_2(\vec{k}_1, \vec{k}_2) \,, \qquad\quad
    \delta_r^{(2)} (\vec{k},a) = \varepsilon\, (D_{1m}^{\rm CDM})^2 \int_{\vec{k}} \mathrm{d}k_{12}F_{2r}(\vec{k}_1, \vec{k}_2)\,,
\end{equation}
with the definition
\begin{align}\label{eq:measure}
&\int_{\vec{k}} \mathrm{d}k_{1 \ldots n}\equiv \int \frac{\mathrm{d}^3 k_1 \ldots \mathrm{d}^3 k_n} {(2\pi)^{3(n-1)}} \,\delta^{(3)}\Big(\vec{k} - \sum_{i\, =\, 1}^n \vec{k}_i \Big)\delta_0 (\vec{k}_1) \ldots \delta_0 (\vec{k}_n)\,. \nonumber
\end{align}
The second order kernels are defined as
\begin{equation}\label{eq:F2_BSM}
\begin{split}
&\widetilde{F}_{2} (\vec{k}_1, \vec{k}_2 ) = [F_2 + \varepsilon f_\chi \Delta F_2](\vec{k}_1, \vec{k}_2 ) =  \frac{5}{7} - \frac{8}{49} \varepsilon f_\chi +\frac{\vec{k}_1 \cdot \vec{k}_2}{2} \left( \frac{1}{k_1^2} + \frac{1}{k_2^2} \right) + \left(\frac{2}{7} + \frac{8}{49} \varepsilon f_\chi \right)  \frac{(\vec{k}_1 \cdot \vec{k}_2)^2}{k_1^2 k_2^2}\,,
\end{split}
\end{equation}
where $a \gg a_{\rm eq}$ was assumed, and 
\begin{align}\label{eq:F2r}
F_{2r} (\vec{k}_1, \vec{k}_2 ) =\,&  \frac{59}{30}  + \frac{17}{6} \frac{\vec{k}_1 \cdot \vec{k}_2}{2} \left( \frac{1}{k_1^2} + \frac{1}{k_2^2} \right) + \frac{13}{15} \frac{(\vec{k}_1 \cdot \vec{k}_2)^2}{k_1^2 k_2^2} \,.
\end{align}
For the velocity divergences we find
\begin{equation}
    \theta_m^{(2)} (\vec{k}, a) =  - f_m \mathcal{H} (D_{1m})^2 \int_{\vec{k}} \mathrm{d}k_{12}\widetilde{G}_2(\vec{k}_1, \vec{k}_2) \,, \qquad\quad 
    \theta_r^{(2)} (\vec{k}, a) = - \varepsilon\, \mathcal{H}  (D_{1m}^{\rm CDM})^2 \int_{\vec{k}} \mathrm{d}k_{12}G_{2r}(\vec{k}_1, \vec{k}_2) \,,
\end{equation}
where
\begin{equation}\label{eq:G2_BSM}
\begin{split}
&\widetilde{G}_{2} (\vec{k}_1, \vec{k}_2 ) =  \frac{3}{7} -  \frac{16}{49} \varepsilon f_\chi + \frac{\vec{k}_1 \cdot \vec{k}_2}{2} \left( \frac{1}{k_1^2} + \frac{1}{k_2^2} \right) + \left(\frac{4}{7} + \frac{16}{49} \varepsilon f_\chi \right)  \frac{(\vec{k}_1 \cdot \vec{k}_2)^2}{k_1^2 k_2^2}\,,
\end{split}
\end{equation}
and
\begin{align}
G_{2r} (\vec{k}_1, \vec{k}_2 ) =\,&  \frac{3}{5}  + \frac{7}{3} \frac{\vec{k}_1 \cdot \vec{k}_2}{2} \left( \frac{1}{k_1^2} + \frac{1}{k_2^2} \right) + \frac{26}{15} \frac{(\vec{k}_1 \cdot \vec{k}_2)^2}{k_1^2 k_2^2} \,.
\end{align}
Following standard convention~\cite{Bernardeau:2001qr} we call $F_2$ and $G_2$ the CDM kernels, namely $\big(\widetilde{F}_2, \widetilde{G}_2\big)\big|_{\varepsilon\, =\, 0} = (F_2, G_2)$. The second order galaxy perturbation is expressed as
\begin{equation}
\delta_g^{(2)} (\vec{k},a) = ( D_{1m} )^2 \int_{\vec{k}} \mathrm{d}k_{12}\Big( F_{2,g} (\vec{k}_1, \vec{k}_2 ) + \varepsilon  F_{2r,g} (\vec{k}_1, \vec{k}_2 )  \Big )\,,
\end{equation}
where
\begin{align}
F_{2,g} (\vec{k}_1, \vec{k}_2 ) =\;& b_1 F_2 (\vec{k}_1, \vec{k}_2) + \frac{b_2}{2} + b_{K^2} \Big(\frac{(\vec{k}_1 \cdot \vec{k}_2)^2}{k_1^2 k_2^2} - \frac{1}{3} \Big)\,,\nonumber \\
F_{2r,g} (\vec{k}_1, \vec{k}_2 ) =\;& b_r F_{2r} (\vec{k}_1, \vec{k}_2 ) \,-\, b_1 \frac{8f_\chi}{49} \Big(1 - \frac{(\vec{k}_1 \cdot \vec{k}_2)^2}{k_1^2 k_2^2} \Big) + \frac{5}{3}b_{mr} + \frac{5}{3}b_{Kr} \Big(\frac{(\vec{k}_1 \cdot \vec{k}_2)^2}{k_1^2 k_2^2} - \frac{1}{3} \Big) \\ &- \mathcal{H}\bigg[b_\theta  G_{2r} (\vec{k}_1, \vec{k}_2) + \frac{5}{3}b_{\nabla\delta} \frac{\vec{k}_1 \cdot \vec{k}_2}{2} \Big( \frac{1}{k_1^2} + \frac{1}{k_2^2} \Big) + \frac{5}{3}b_{\delta\theta} + \frac{5}{3} b_K \Big(\frac{(\vec{k}_1 \cdot \vec{k}_2)^2}{k_1^2 k_2^2} - \frac{1}{3} \Big) \bigg] \,. \nonumber
\end{align}
Notice that, up to the second order, the bias coefficients $b_{mr}$ and $b_{\delta\theta}$ could be absorbed into a redefinition of $b_2$, whereas $b_{Kr}$ and $b_K$ could be absorbed into a redefinition of $b_{K^2}$. The galaxy bispectrum
\begin{equation}
    \langle \delta_g  (\vec{p}) \delta_g (\vec{p}_1) \delta_g (\vec{p}_2) \rangle = (2\pi)^3 \delta^{(3)} (\vec{p} + \vec{p}_1 + \vec{p}_2 ) B_{g} (\vec{p}, \vec{p}_1, \vec{p}_2 )
\end{equation}
is found to be at tree level
\begin{align}
  B_{g} (\vec{p}, \vec{p}_1, \vec{p}_2 ) =&\, \left[ 1 + 4\varepsilon f_\chi \Big( \log \frac{a}{a_{\rm eq}} - \frac{7}{3}\Big)   \right] B_{g}^{\rm CDM} (\vec{p}, \vec{p}_1, \vec{p}_2 )  \label{eq:bispectrum} \\
  +&\, 2 b_1 \varepsilon   \bigg\{P_{m, L}^{\rm CDM} (p)  P_{m, L}^{\rm CDM} (p_1) \bigg[  b_1 F_{2r, g} (\vec{p}, \vec{p}_1 ) + \frac{10\hspace{0.4mm} \widehat{b}_r}{3}  F_{2, g} (\vec{p}, \vec{p}_1 ) \bigg] + \mathrm{permutations} \bigg\}\,, \nonumber 
\end{align}
where
\begin{equation}\label{eq:Bg_CDM}
B_{g}^{\rm CDM} (\vec{p}, \vec{p}_1, \vec{p}_2 ) = 2 b_1^2 P_{m, L}^{\rm CDM} (p)  P_{m, L}^{\rm CDM} (p_1) F_{2,g}(\vec{p}, \vec{p}_1) + \mathrm{permutations}
\end{equation}
is the standard expression. Taking the squeezed limit of~\eqref{eq:bispectrum} we find
\begin{align}
\frac{B_g (\vec{p}, \vec{k} - \vec{p}/2, - \vec{k} - \vec{p}/2 )}{P_{m, L}(p)} \bigg|_{\vec{p}\, \to \, 0} = -\, b_1^3 \frac{(\vec{p}\cdot \vec{k})^2}{p^2 k}\, \frac{\partial P_{m, L}(k)}{\partial k} \label{eq:BAO_B} \bigg[ 1 + \frac{\varepsilon}{b_1} \Big( \frac{37}{6} b_r - \frac{5}{3} \mathcal{H} \Big(b_{\nabla \delta} + \frac{17}{5} b_\theta \Big) \Big) \bigg] + \mathcal{O} \big( P_{m, L}(k) \big)\,.
 \end{align}
In a $\Lambda$CDM model the coefficient in square parentheses is fixed to $1$ by the equivalence principle (see {\it e.g.} Ref.~\cite{Baldauf:2015xfa}), hence this represents a genuine violation of the consistency relations~\cite{Lewandowski:2019txi}. Analytical and numerical estimates of relative bias parameters~\cite{Schmidt:2016coo,Barreira:2019qdl} indicate that in a $\Lambda$CDM model $b_r , \mathcal{H} b_\theta \sim \mathcal{O}(1) $, while $b_{\nabla \delta}$ could be suppressed compared to the other two.

For two different tracers $A$ and $B$, the $1/p$ pole of the squeezed bispectrum no longer vanishes
\begin{equation}
B^{AAB}_g (\vec{p}, \vec{p}_1, \vec{p}_2 )\Big|_{\vec{p}\, \to \, 0} \simeq \varepsilon P_{m, L}^{\rm CDM} (p) P_{m, L}^{\rm CDM} (p_1)\frac{\vec{p}\cdot \vec{p}_1}{p^2}  \frac{7}{6} b_1^A \bigg\{ b_1^A \Big[ b_r^B - \frac{10}{7}\mathcal{H} \Big(b_{\nabla\delta}^B + \frac{2}{5}b_\theta^B \Big) \Big] - (A \leftrightarrow B) \bigg\}\,,  \label{eq:pole_B_real_bis}
\end{equation}
which is maximized by picking tracers that are as different as possible. The pole is sensitive to a different linear combination of the relative bias parameters that already appeared in~\eqref{eq:BAO_B}.

The bias expansion in~\eqref{eq:bias_second} can be straightforwardly extended to third order in the fields. As long as we are concerned with the 1-loop power spectrum including log-enhanced new physics corrections, it is sufficient to consider only one additional bias parameter. We choose this to be the coefficient of the following operator~\cite{McdonaldRoy,Chan:2012jj,Assassi:2014fva},
\begin{equation}
\delta_g \supset b_{\Gamma_3} \left[ \left( \frac{\nabla_i \nabla_j}{\nabla^2} \delta_m \right)^2 - \delta_m^2 -  \left( \frac{\nabla_i \nabla_j}{\nabla^2} \frac{\theta_m}{f_m \mathcal{H}} \right)^2 + \frac{\theta_m^2}{f_m^2 \mathcal{H}^2} \right]\,,
\end{equation}
which starts at third order in PT.\footnote{\texttt{PyBird}~\cite{DAmico:2020kxu,PyBird} uses a different basis for the nonlinear bias coefficients, related to ours up to third order by $(b_1)_{\rm P} = b_1$, $(b_2)_{\rm P} = b_1 + 7 b_{K^2}/2$ and $(b_4)_{\rm P} = b_2/2 - 17 b_{K^2}/6$, whereas $(b_3)_{\rm P} = b_1 + 15 b_{K^2} + 6\hspace{0.3mm} b_{\Gamma_3}$. Relevant linear combinations are $\sqrt{2}\,(c_2)_{\mathrm{P}} \equiv (b_2 + b_4)_{\rm P} = b_1 + b_2/2 + 2 b_{K^2}/3$ and $\sqrt{2}\,(c_4)_{\mathrm{P}} \equiv (b_2 - b_4)_{\rm P} = b_1 - b_2/2 + 19 b_{K^2}/3$. Yet another basis is used by \texttt{CLASS-PT}~\cite{Chudaykin:2020aoj}, with $(b_1)_{\rm C} = b_1$, $(b_2)_{\rm C} = b_2 + 4 b_{K^2}/3$ and $(b_{\mathcal{G}_2})_{\rm C}= b_{K^2}$, while $(b_{\Gamma_3})_{\rm C} = b_{\Gamma_3}$.\label{foot:bias}} In addition, among the many possible operators involving relative perturbations we include~\cite{Schmidt:2016coo}
\begin{equation}
\delta_g \supset b_{\rm nloc}^r \big(\nabla^i v_r^j - \delta^{ij} \theta_r \big) \frac{\nabla_i \nabla_j}{\nabla^2} \left[ \delta_m^2 - \left(\frac{\nabla_a \nabla_b}{\nabla^2} \delta_m\right)^2 \right]\,. 
\end{equation}
The third-order solutions for the total and relative density contrasts are
\begin{equation}
    \delta_m^{(3)} (\vec{k}, a) =  (D_{1m})^3 \int_{\vec{k}} \mathrm{d}k_{123}\widetilde{F}_3(\vec{k}_1, \vec{k}_2, \vec{k}_3) \,, \qquad
    \delta_r^{(3)} (\vec{k}, a) = \varepsilon (D_{1m}^{\rm CDM})^3 \int_{\vec{k}} \mathrm{d}k_{123}F_{3r}(\vec{k}_1, \vec{k}_2, \vec{k}_3)\,,
\end{equation}
where $\widetilde{F}_{3} (\vec{k}_1, \vec{k}_2, \vec{k}_3) = [F_{3} + \varepsilon f_\chi \Delta F_{3}] (\vec{k}_1, \vec{k}_2, \vec{k}_3)  = \tfrac{1}{3!}\sum_{\rm perm} \big[ P_{3}(\vec{k}_1, \vec{k}_2, \vec{k}_3) +  \varepsilon f_\chi \Delta P_3 (\vec{k}_1, \vec{k}_2, \vec{k}_3) \big]$, with the standard $P_3$ defined in Eq.~(A.3) of Ref.~\cite{Goroff:1986ep} and 
\begin{align}
&\Delta P_3 (\vec{k}_1, \vec{k}_2, \vec{k}_3) = \frac{4}{3969\, k_1^2 k_2^2 k_3^2 k_{23}^2}\bigg\{(\vec{k}_1 \cdot \vec{k}_{23})^2 \big[ k_2^2 (49\, \vec{k}_2 \cdot \vec{k}_3 - 30\,k_3^2) + \vec{k}_2 \cdot \vec{k}_3 (128\, \vec{k}_2 \cdot \vec{k}_3 + 49\, k_3^2 ) \big]  
\nonumber \\ &+ \vec{k}_1 \cdot \vec{k}_{23} \big[ (\vec{k}_2 \cdot \vec{k}_3)^2 - k_2^2 k_3^2 \big] (113\, k_{23}^2 + 162\, k_1^2 ) - 49\, k_1^2 k_{23}^2 \big[ \vec{k}_2 \cdot \vec{k}_3 (k_3^2 - 3\, \vec{k}_2 \cdot \vec{k}_3 ) + k_2^2 (\vec{k}_2 \cdot \vec{k}_3 + 5\, k_3^2 )  \big]\bigg\}\,,
\end{align}
where $\vec{k}_{ij} \equiv \vec{k}_i + \vec{k}_j$. For the relative density perturbation we find $F_{3r} (\vec{k}_1, \vec{k}_2, \vec{k}_3) = \tfrac{1}{3!}\sum_{\rm perm} P_{3r} (\vec{k}_1, \vec{k}_2, \vec{k}_3)$ with
\begin{align}
&P_{3r}(\vec{k}_1, \vec{k}_2, \vec{k}_3) = \frac{1}{8820\, k_1^2 k_2^2 k_3^2 k_{12}^2 k_{23}^2} \bigg\{k_{12}^2 \vec{k}_1 \cdot \vec{k}_{23} \bigg[4\, \vec{k}_1 \cdot \vec{k}_{23}\Big(k_2^2 (595\, \vec{k}_{2}\cdot \vec{k}_3 + 342\, k_3^2) + \vec{k}_2 \cdot \vec{k}_3 (848\, \vec{k}_2 \cdot \vec{k}_3 + 595\, k_3^2)\Big) \nonumber  \\
+&\; 3k_1^2 \Big(k_2^2 (1645\, \vec{k}_2 \cdot \vec{k}_3 + 906\, k_3^2) + \vec{k}_2 \cdot \vec{k}_3 (2384\, \vec{k}_2 \cdot \vec{k}_3 + 1645\, k_3^2 ) \Big)\bigg] + k_{23}^2 \bigg[ 50\, \vec{k}_1 \cdot \vec{k}_2 \bigg( \vec{k}_{12} \cdot \vec{k}_3 \Big(4\,\vec{k}_{12} \cdot \vec{k}_3(8\,\vec{k}_1 \cdot \vec{k}_2 + 7\, k_2^2) \nonumber \\ 
+&\; k_{12}^2 (44\, \vec{k}_1 \cdot \vec{k}_2 +63\, k_2^2 )  \Big) + k_3^2 \Big(9\,\vec{k}_{12} \cdot \vec{k}_3 (8\, \vec{k}_1\cdot \vec{k}_2 + 7\, k_2^2 ) + 14k_{12}^2 (6\, \vec{k}_1 \cdot\vec{k}_2 + 7\, k_2^2) \Big)\bigg) + 14\, k_{12}^2 \vec{k}_1 \cdot \vec{k}_{23} \Big( \vec{k}_2 \cdot \vec{k}_3 (316\, \vec{k}_2 \cdot \vec{k}_3 \nonumber \\ 
+&\; 405\, k_3^2 ) + k_2^2 (405\, \vec{k}_2 \cdot \vec{k}_3 + 494\, k_3^2 ) \Big)  + k_1^2 \bigg( 8184 (\vec{k}_2 \cdot \vec{k}_3)^2 k_{12}^2 + 350\, \vec{k}_1 \cdot \vec{k}_2\, \vec{k}_{12}\cdot \vec{k}_3 (4\, \vec{k}_{12} \cdot \vec{k}_3 + 9\, k_{12}^2)  \nonumber\\ +&\; 175\, k_3^2 \Big(k_{12}^2 (28\, \vec{k}_1 \cdot \vec{k}_2 + 47\, \vec{k}_2 \cdot \vec{k}_3) + 18\, \vec{k}_1 \cdot \vec{k}_2\, \vec{k}_{12}\cdot \vec{k}_3 \Big) + k_2^2 \Big(8225\, \vec{k}_2 \cdot \vec{k}_3 k_{12}^2 + 2\, k_{12}^2 (2050\, \vec{k}_{12} \cdot \vec{k}_3 + 6933\, k_3^2) \nonumber \\ 
+&\; 300\, \vec{k}_{12}\cdot \vec{k}_3 (4\, \vec{k}_{12}\cdot \vec{k}_3 + 9\, k_3^2) \Big)\bigg)\bigg] \bigg\}\,.
\end{align}
For the velocity divergences at third order we obtain
\begin{equation}
\theta_m^{(3)} (\vec{k}, a) = - f_m \mathcal{H} (D_{1m})^3 \int_{\vec{k}} \mathrm{d}k_{123}\widetilde{G}_3(\vec{k}_1, \vec{k}_2, \vec{k}_3)\,, \qquad\quad
\theta_r^{(3)} (\vec{k}, a) = -\, \varepsilon \mathcal{H} (D_{1m}^{\rm CDM})^3 \int_{\vec{k}} \mathrm{d}k_{123} G_{3r}(\vec{k}_1, \vec{k}_2, \vec{k}_3)\,,
\end{equation}
where
\begin{align}
\widetilde{G}_{3}(\vec{k}_1, \vec{k}_2, \vec{k}_3) =&\; 3\, \widetilde{F}_{3} (\vec{k}_1, \vec{k}_2, \vec{k}_3) - \frac{1}{3!}\sum_{\rm perm} \left[  \left( 1 + \frac{\vec{k}_1 \cdot \vec{k}_{23}}{k_{1}^2} \right) \widetilde{F}_2 (\vec{k}_2, \vec{k}_3) + \left( 1 + \frac{\vec{k}_1 \cdot \vec{k}_{23}}{k_{23}^2} \right) \widetilde{G}_2 (\vec{k}_2, \vec{k}_3) \right]\,, \nonumber \\
G_{3r}(\vec{k}_1, \vec{k}_2, \vec{k}_3) =&\; 3\, F_{3r} (\vec{k}_1, \vec{k}_2, \vec{k}_3) - \frac{1}{3!}\sum_{\rm perm} \bigg[  \left( 1 + \frac{\vec{k}_1 \cdot \vec{k}_{23}}{k_{1}^2} \right) \left( F_{2r} (\vec{k}_2, \vec{k}_3) + \frac{5}{3} F_2 (\vec{k}_2, \vec{k}_3)\right) \\ &\qquad\qquad\qquad\qquad\qquad\qquad\qquad +  \left( 1 + \frac{\vec{k}_1 \cdot \vec{k}_{23}}{k_{23}^2} \right) \left( G_{2r} (\vec{k}_2, \vec{k}_3) + \frac{5}{3} G_2 (\vec{k}_2, \vec{k}_3) \right) \bigg]\, . \nonumber
\end{align}
The IR behavior is determined by the following shift terms
\begin{equation}\label{eq:F3tilde_IR}
\widetilde{F}_3, F_{3r}, \widetilde{G}_3, G_{3r} \supset \frac{C_{\widetilde{F}_3}, C_{F_{3r}}, C_{\widetilde{G}_3}, C_{G_{3r}}}{6} \bigg( \frac{\vec{k}_1 \cdot \vec{k}_3}{k_1^2}\frac{\vec{k}_2 \cdot \vec{k}_3}{k_2^2}\; + \frac{\vec{k}_1 \cdot \vec{k}_2}{k_1^2}\frac{\vec{k}_3 \cdot \vec{k}_2}{k_3^2} + \frac{\vec{k}_2 \cdot \vec{k}_1}{k_2^2}\frac{\vec{k}_3 \cdot \vec{k}_1}{k_3^2} \bigg)\,,
\end{equation}
where $C_{\widetilde{F}_3} = C_{\widetilde{G}_3} = 1$ are not modified relative to CDM up to $\mathcal{O}(\beta)$, similarly to the second terms of $\widetilde{F}_2$ and $\widetilde{G}_2$ in Eqs.~(\ref{eq:F2_BSM}) and~(\ref{eq:G2_BSM}), respectively. For the relative density perturbation we find $C_{F_{3r}} = 4$ and $C_{G_{3r}} = 3$.

The third order galaxy perturbation is expressed as
\begin{equation}
\delta_g^{(3)} (\vec{k},a) = ( D_{1m} )^3 \int_{\vec{k}} \mathrm{d}k_{123}\Big( F_{3,g} (\vec{k}_1, \vec{k}_2, \vec{k}_3 ) + \varepsilon  F_{3r,g} (\vec{k}_1, \vec{k}_2, \vec{k}_3 )  \Big )\,,
\end{equation}
where the standard kernel is
\begin{align}
F_{3,g} (\vec{k}_1, \vec{k}_2, \vec{k}_3 ) = b_1 F_3 (\vec{k}_1, \vec{k}_2, \vec{k}_3 ) + \frac{1}{3!}\sum_{\rm perm} \bigg[ b_2 F_2 (\vec{k}_2, \vec{k}_3) +&\, 2 b_{K^2} \left( \frac{(\vec{k}_1 \cdot \vec{k}_{23})^2}{k_1^2 k_{23}^2} - \frac{1}{3} \right) F_2 (\vec{k}_2, \vec{k}_3) \\
+&\, \frac{4}{7} b_{\Gamma_3} \left( \frac{(\vec{k}_1 \cdot \vec{k}_{23})^2}{k_1^2 k_{23}^2} - 1 \right)\left( 1 - \frac{(\vec{k}_2 \cdot \vec{k}_3)^2}{k_2^2 k_3^2} \right)\bigg],\nonumber
\end{align}
while the $\mathcal{O}(\varepsilon)$ correction reads
\begin{align}
&\,F_{3r,g} (\vec{k}_1, \vec{k}_2, \vec{k}_3 ) = [b_1 f_\chi \Delta F_3 + b_r F_{3r}] (\vec{k}_1, \vec{k}_2, \vec{k}_3 ) + \frac{1}{3!}\sum_{\rm perm} \bigg\{ \bigg[ b_2 + 2 b_{K^2} \left( \frac{(\vec{k}_1 \cdot \vec{k}_{23})^2}{k_1^2 k_{23}^2} - \frac{1}{3} \right) - 2 b_{\Gamma_3} \left( \frac{(\vec{k}_1 \cdot \vec{k}_{23})^2}{k_1^2 k_{23}^2} - 1 \right)\bigg] \nonumber  \\
&\,\times \frac{8f_\chi}{49} \left( \frac{(\vec{k}_2 \cdot \vec{k}_3)^2}{k_2^2 k_3^2}  - 1 \right) + \bigg(b_{mr} + b_{Kr} \left( \frac{(\vec{k}_1 \cdot \vec{k}_{23})^2}{k_1^2 k_{23}^2} - \frac{1}{3} \right)  \bigg) \Big[F_{2r} + \frac{5}{3} F_{2}\Big](\vec{k}_2, \vec{k}_3) \bigg\} - \mathcal{H} b_\theta G_{3r}(\vec{k}_1, \vec{k}_2, \vec{k}_3 )  \nonumber \\
&\,\hspace{1.0cm}- \mathcal{H}\, \frac{1}{3!}\sum_{\rm perm} \bigg\{b_{\delta\theta} \Big[G_{2r} + \frac{5}{3}F_2\Big](\vec{k}_2, \vec{k}_3) + b_{\nabla\delta} \left(\frac{\vec{k}_1 \cdot \vec{k}_{23}}{k_{23}^2} G_{2r}(\vec{k}_2, \vec{k}_3) + \frac{5}{3} \frac{\vec{k}_1 \cdot \vec{k}_{23}}{k_{1}^2} F_{2}(\vec{k}_2, \vec{k}_3) \right) \nonumber \\ &\,\hspace{1.5cm}+ b_K \left( \frac{(\vec{k}_1 \cdot \vec{k}_{23})^2}{k_1^2 k_{23}^2} - \frac{1}{3} \right) \Big[G_{2r} + \frac{5}{3} F_2 \Big](\vec{k}_2, \vec{k}_3) + \frac{5}{3} b_{\rm nloc}^r \left( \frac{(\vec{k}_1 \cdot \vec{k}_{23})^2}{k_1^2 k_{23}^2} - 1 \right) \left( 1 - \frac{(\vec{k}_2 \cdot \vec{k}_3)^2}{k_2^2 k_3^2} \right)\bigg\}\,.
\end{align}
We see that, up to third order, the coefficient $b_{\rm nloc}^r$ can be absorbed into a redefinition of $b_{\Gamma_3}$.

So far, we have not discussed whether the new dynamics in the dark sector studied in this work requires new counterterms in the EFT. This can be easily checked by taking the ultraviolet (UV) limit of the 1-loop terms contributing to the galaxy power spectrum. Under the assumption of a negligible difference between baryons and DM at $\mathcal{O}(\varepsilon^0)$, we find that the standard leading order counterterm proportional to $k^2 P_{m,L}(k)$ is sufficient to absorb all the UV sensitivities. Notice that in general an additional counterterm proportional to $k^2 P_{mr,L}(k)$ is required, if the baryons and DM have different transfer functions~\cite{Lewandowski:2014rca}. This is actually the case in our model, since the new force produces scale dependence in the DM clustering at $\mathcal{O}(\varepsilon)$~\cite{Archidiacono:2022iuu}. However, this effect is neglected here because it is not log enhanced. A genuine new counterterm is instead required for $P_{mr}^{(13)}$, whose dependence on the UV cutoff $\Lambda$ reads
\begin{align}
    P_{mr}^{(13)} (k)\;\xrightarrow{\rm UV}\; - 
    \frac{P_{mr, L} (k)} { 30 } \int_0^\Lambda \frac{\mathrm{d} p}{2 \pi^2 } \, p^2 P_{m, L} (p)  \equiv - 
    \frac{P_{mr, L} (k)} { 30 } \sigma^2_\Lambda \,,
\end{align}
where $P_{mr, L}(k) = 5\varepsilon P_{m,L}^{\rm CDM}(k)/3$. The variance of the density field $\sigma^2_\Lambda$ is highly sensitive to the amplitude of the small scale modes, well beyond the perturbative regime. Its effect must therefore be removed by adding a suitable counterterm, $\alpha(\tau) P_{mr, L}(k)$, to the 1-loop galaxy power spectrum. 
At linear order, this new term is degenerate with the bias parameters $b_{r}$ or $b_\theta$, whose value could thus change by $\mathcal{O}(1)$ compared to a naive peak-background-split estimate. Our results are in agreement with Ref.~\cite{Braganca:2020nhv}, where the baryon-DM two fluid system was discussed in a standard cosmological model. In conclusion, for the log-enhanced corrections studied in this work it is sufficient to include the standard counterterm $k^2 P_{m,L} (k)$. 

\subsection*{Redshift space}
Up to second order in the fields and $\mathcal{O}(\beta)$, the velocity bias expansion is~\cite{Schmidt:2016coo}
\begin{equation}
v_g^i = v_m^i + c_r v_r^i + c_{mr} \delta_m v_r^i + c_K K^{ij} v_{r j}\,,
\end{equation}
where the coefficient of $v_m^i$ is fixed to $1$ by Galilean invariance. This yields $\theta_g^{(1)}(\vec{k}, a) = - f_m \mathcal{H} D_{1m} \big(1 + \varepsilon\hspace{0.4mm} 5 c_r /3\big) \delta_0(\vec{k})$ at linear order, while at second order
\begin{equation}
\theta_g^{(2)} (\vec{k}, a ) = - f_m \mathcal{H} (D_{1m})^2  \int_{\vec{k}} \mathrm{d}k_{12} \Big( G_2 (\vec{k}_1, \vec{k}_2 ) + \varepsilon G_{2r, g} (\vec{k}_1 , \vec{k}_2 ) \Big)\,,    
\end{equation}
with
\begin{align}
G_{2r, g} (\vec{k}_1 , \vec{k}_2 ) = c_r G_{2r} (\vec{k}_1, \vec{k}_2) + \frac{5}{3} \bigg[ c_{mr} - \frac{c_K}{3} +  \Big(c_{mr}   + \frac{2}{3}c_K\Big) \frac{\vec{k}_1 \cdot \vec{k}_2}{2} \Big( \frac{1}{k_1^2} + \frac{1}{k_2^2} \Big)  &\,+ c_K \frac{(\vec{k}_1 \cdot \vec{k}_2)^2}{k_1^2 k_2^2} \bigg] \\ 
&\,- \frac{16f_\chi}{49}  \Big( 1 - \frac{(\vec{k}_1 \cdot \vec{k}_2)^2}{k_1^2 k_2^2} \Big) \,.     \nonumber
\end{align}
The linear galaxy density perturbation including Redshift Space Distortions (RSD) then reads
\begin{equation}
\delta_{g,\, \mathrm{RSD}}^{(1)}(\vec{k}, a) = D_{1m} \Big[ b_1 + f_m \mu_k^2 + \frac{5}{3} \varepsilon\big(\, \widehat{b}_r + c_r f_m \mu_k^2 \big) \Big] \delta_0 (\vec{k})\,,
\end{equation}
where $\mu_k \equiv \vec{k} \cdot \widehat{\vec{z}}/k$ is the cosine of the angle between the wavevector and the line of sight, the latter taken along the $z$ axis. At second order
\begin{equation}
\delta_{g, \mathrm{RSD}}^{(2)}(\vec{k}, a)\,= ( D_{1m} )^2\int_{\vec{k}} \mathrm{d}k_{12}\Big(F_{2, g}^{\rm RSD}(\vec{k}_1, \vec{k}_2) + \varepsilon F_{2r, g}^{\rm RSD}(\vec{k}_1, \vec{k}_2 ) \Big)\,,
\end{equation}
where
\begin{align}
F_{2, g}^{\rm RSD}(\vec{k}_1, \vec{k}_2) =&\, F_{2,g}(\vec{k}_1, \vec{k}_2) + f_m \mu_k^2 G_2 (\vec{k}_1, \vec{k}_2) + b_1 f_m \mu_k k  \frac{\mu_{k_1}k_2 + \mu_{k_2}k_1}{2k_1 k_2} + f_m^2 \mu_k^2 k^2  \frac{\mu_{k_1}\mu_{k_2}}{2k_1 k_2}\,, \\
F_{2r, g}^{\rm RSD}(\vec{k}_1, \vec{k}_2 ) =&\, F_{2r, g}(\vec{k}_1, \vec{k}_2) + \mu_k^2 G_{2r, g}(\vec{k}_1, \vec{k}_2) + \frac{5}{3} f_m  \mu_k k\big(  c_r b_1 + \widehat{b}_r  \big) \frac{\mu_{k_1} k_2 + \mu_{k_2}k_1}{2k_1 k_2} +  \frac{10\hspace{0.2mm}c_r}{3}f_m^2 \mu_k^2 k^2\, \frac{\mu_{k_1}\mu_{k_2}}{2k_1 k_2} \,, \nonumber
\end{align}
with $\vec{k} = \vec{k}_1 + \vec{k}_2\,$. The above expressions allow one to obtain the redshift space expression of the tree-level galaxy bispectrum, generalizing Eqs.~(\ref{eq:bispectrum}) and~(\ref{eq:Bg_CDM}). The above calculations can also be straightforwardly extended to the third perturbative order, to evaluate the galaxy power spectrum at one loop.

\section{Lagrangian perturbation theory}\label{sec:lagrangian}

From the geodesic equation for $\chi$~\cite{Archidiacono:2022iuu} we derive the Lagrangian Perturbation Theory (LPT) equation obeyed by the displacement, defined by $\vec{x} = \vec{q} + \bm{\psi}(\vec{q}, \tau)$,
\begin{equation}
\bm{\psi}_{\chi}^{\prime\prime} +     (\mathcal{H} + \widetilde{m}_s s') \bm{\psi}_{\chi}^{\prime} + \bm{\nabla}_{\chi} \big( \Psi (\vec{x}_\chi) + \widetilde{m}_s  \delta s \big) = 0\,,
\end{equation}
where $\nabla_\chi^i = \partial/\partial x_{\chi i}\,$. The equation for the baryons is simply
\begin{equation}
\bm{\psi}_{b}^{\prime\prime} +     \mathcal{H} \bm{\psi}_{b}^{\prime} + \bm{\nabla}_{b}  \Psi (\vec{x}_b) = 0\,.   
\end{equation}
Switching variables to the total and relative displacements and working up to $\mathcal{O}(\beta)$ and at second order in the fields, we rewrite the equations as
\begin{align}
&\bm{\psi}_m^{\prime\prime} + \mathcal{H}(1 - f_\chi \varepsilon ) \bm{\psi}^\prime_m = - \bm{\nabla}_m \Psi (\vec{x}_m) - f_\chi \widetilde{m}_s \bm{\nabla}_\chi \delta s (\vec{x}_\chi)\,, \nonumber \\    
&\bm{\psi}_r^{\prime\prime} + \mathcal{H} \bm{\psi}_r^{\prime} = - \bm{\psi}_r \cdot \bm{\nabla} (\bm{\nabla} \Psi) - \widetilde{m}_s \bm{\nabla}_\chi \delta s (\vec{x}_\chi) + \mathcal{H} \varepsilon \bm{\psi}_m^\prime\,,
\end{align}
with $\nabla^i = \partial/\partial q_i\,$. Taking the sub-horizon limit, they are solved at linear level by
\begin{equation}
\bm{\psi}_{m,r}^{(1)} (\vec{k}, a)  = \frac{i \vec{k}}{k^2} D_{1m, 1r} \delta_0 (\vec{k}) = \frac{i \vec{k}}{k^2} \delta_{m,r}^{(1)}(\vec{k}, a)\,,    
\end{equation}
where the last equality follows from~\eqref{eq:linear_SPT}. At second order we find
\begin{equation}
\bm{\psi}_m^{(2)} (\vec{k}, a) = - \frac{i \vec{k}}{2k^2} D_{2m}\int_{\vec{k}} \mathrm{d}k_{12} \bigg( 1 - \frac{(\vec{k}_1 \cdot \vec{k}_2)^2}{k_1^2 k_2^2} \bigg)\,,    
\end{equation}
where the second order growth factor reads for $a \gg a_{\rm eq}$
\begin{equation}
D_{2m} \simeq - \frac{3}{7} (D_{1m})^2 + \frac{16}{49} f_\chi \varepsilon (D_{1m}^{\rm CDM})^2\,.
\end{equation}
For the relative displacement we find
\begin{equation}
\bm{\psi}_r^{(2)} (\vec{k},a) = - \frac{i \vec{k}}{2k^2} \int_{\vec{k}} \mathrm{d}k_{12} \bigg[ D_{2r}^{\rm np} \bigg(1 - \frac{(\vec{k}_1 \cdot \vec{k}_2)^2}{k_1^2 k_2^2} \bigg) + D_{2r}^{\rm p} \bigg(1 + \frac{\vec{k}_1 \cdot \vec{k}_2}{2}\Big(\frac{1}{k_1^2} + \frac{1}{k_2^2} \Big) \bigg) \bigg]\,,
\end{equation}
where
\begin{equation}
D_{2r}^{\rm p}= -\frac{5}{8} D_{2r}^{\rm np} \simeq \varepsilon (D_{1m}^{\rm CDM})^2 \,.    
\end{equation}
The subscripts ``p'' and ``np'' stand for ``pole'' and ``no pole'', referring to presence or absence of a divergence for $\vec{k}_1 \ll \vec{k}_2$ in the associated form factor. These LPT solutions are related to the SPT ones presented in Appendix~\ref{sec:eulerian} as follows. For the total matter perturbation we have
\begin{equation}
\delta_m^{(2)}(\vec{q}, a)= - \nabla_i \psi_m^{(2)i} + \frac{1}{2}(\nabla_i \psi_m^{(1)i})^2 + \frac{1}{2}\nabla_i \psi_m^{(1)j} \nabla_j \psi_m^{(1)i} + \nabla_i \nabla_j \psi_m^{(1)j} \cdot \psi_m^{(1)i}\,,    
\end{equation}
or in Fourier space
\begin{equation}
\delta_m^{(2)}(\vec{k}, a) = (D_{1m})^2 \int_{\vec{k}} \mathrm{d}k_{12} \bigg[\frac{1}{2}\bigg(1 - \frac{D_{2m}}{D_{1m}^2}\bigg) + \frac{1}{2}\bigg(1 + \frac{D_{2m}}{D_{1m}^2}\bigg) \frac{(\vec{k}_1 \cdot \vec{k}_2)^2}{k_1^2 k_2^2} + \frac{\vec{k}_1 \cdot \vec{k}_2}{2} \Big( \frac{1}{k_1^2} + \frac{1}{k_2^2} \Big)  \bigg]\,.
\end{equation}
Plugging in the explicit forms of the growth factors, the terms proportional to $\varepsilon \log a/a_{\rm eq}$ cancel out, giving agreement with Eqs.~(\ref{eq:second_order}) and~(\ref{eq:F2_BSM}). For the relative density perturbation we obtain
\begin{equation}
\delta_r^{(2)}(\vec{q}, a) = - \nabla_i \psi_r^{(2)i} + \nabla_i \psi_m^{(1)i} \nabla_j \psi_r^{(1)j} + \nabla_i \psi_m^{(1)j}\nabla_j \psi_r^{(1)i} +\nabla_i \nabla_j \psi_m^{(1)j} \cdot \psi_r^{(1)i} + \nabla_i \nabla_j \psi_r^{(1)j} \cdot \psi_m^{(1)i}\,,
\end{equation}
which in Fourier space reads
\begin{equation}
\delta_r^{(2)}(\vec{k}, a) = 2 D_{1m} D_{1r} \int_{\vec{k}} \mathrm{d}k_{12} \bigg[\frac{1}{2}\bigg(1 - \frac{D_{2r}^{\rm np}+D_{2r}^{\rm p}}{2D_{1m}D_{1r}}\bigg) + \frac{1}{2}\bigg(1 + \frac{D_{2r}^{\rm np}}{2 D_{1m}D_{1r}}\bigg) \frac{(\vec{k}_1 \cdot \vec{k}_2)^2}{k_1^2 k_2^2} + \bigg( 1 - \frac{D_{2r}^{\rm p}}{4D_{1m}D_{1r}} \bigg) \frac{\vec{k}_1 \cdot \vec{k}_2}{2} \Big( \frac{1}{k_1^2} + \frac{1}{k_2^2} \Big)  \bigg]\,.
\end{equation}
This agrees with Eqs.~(\ref{eq:second_order}) and~(\ref{eq:F2r}) once the expressions of the growth factors are used. The above calculations could be extended straightforwardly, if tediously, to the third perturbative order.

\section{Consistency relations}\label{sec:consistency_rel}
In this appendix we provide two arguments, one in SPT and one in LPT, that prove the absence of $\mathcal{O}(\beta)$ bulk flows in the $1$-loop galaxy power spectrum.

In SPT, we start by noticing that in Eqs.~(\ref{eq:F2_BSM}) and~(\ref{eq:G2_BSM}) the pieces linear in $\vec{k}_1 \cdot \vec{k}_2$ (shift terms) are not corrected from their CDM values at $\mathcal{O}(\beta)$. The same applies for any $n$ to the shift terms in $\widetilde{F}_n$ and $\widetilde{G}_n$, which control the IR behavior of the correlators,\footnote{See~\eqref{eq:F3tilde_IR} for the expressions of the shift terms at third order.} because at $\mathcal{O}(\beta)$ the total matter perturbations satisfy the single-fluid equations in~\eqref{eq:tot_pert} at all perturbative orders. As a result, at $\mathcal{O}(\beta)$ the consistency relations are always satisfied by the $m$ fluid alone, and $P_m$ is free from IR divergences at any loop order. However, {\it a priori} an IR sensitivity at $\mathcal{O}(\beta)$ may arise in the cross power spectrum $P_{mr}$. To see that this is, in fact, not the case, we write
\begin{equation}\label{eq:P_mr}
P_{mr} = \frac{1}{2} (P_\chi - P_b) + \mathcal{O}(\beta^2)
\end{equation}
and observe that, separately for each of $b$ and $\chi$, it is possible to remove the effect of the gradient of the effective potential (for $b$ this is of pure gravitational origin, while for $\chi$ it combines the effects of gravity and dark force), by changing coordinates to the free-falling frame. This transformation is at the heart of the derivation of the consistency relations~\cite{Creminelli:2013poa}, which ensure the cancellation of bulk flows (IR finiteness) in the single-species power spectrum. Although the free-falling frames of the two species are different, in the power spectrum the difference only arises at $\mathcal{O}(\beta^2)$, being proportional to the square of the relative velocity $\vec{v}_r$. We thus conclude that, at $\mathcal{O}(\beta)$, $P_{mr}$ must be IR finite. We finally note that from~\eqref{eq:deltam_deltar} follows 
\begin{equation}\label{eq:P_chib}
P_{\chi b} = P_m + (1 - 2 f_\chi) P_{mr} - f_\chi (1 - f_\chi)P_r\,,    
\end{equation}
where $P_{mr}$ and $P_m$ do not contain IR divergences at $\mathcal{O}(\beta)$, while $P_r$ is by construction of order $\beta^2$. As a consequence, $P_{\chi b}$ must also be IR-finite at $\mathcal{O}(\beta)$, completing our argument for the density perturbations.

The galaxy power spectrum receives contributions also from correlators involving the relative velocity, such as for instance $\langle \delta_m \theta_r \rangle$. Similarly to Eq.~(\ref{eq:P_mr}), we write
\begin{equation}\label{eq:IR_vel}
\langle \delta_\chi \theta_\chi  -  \delta_b \theta_b \rangle = \langle \delta_m \theta_r + \theta_m \delta_r \rangle + \mathcal{O}(\beta^2)
\end{equation}
and observe that $\langle \theta_m \delta_r \rangle$ has the same IR structure as $\langle \delta_m \delta_r \rangle =P_{mr}$, because $\theta_m$ and $\delta_m$ have identical shift terms (a result inherited from CDM). Since the left-hand side of Eq.~(\ref{eq:IR_vel}) contains single-species correlators, which are IR finite, by combining the previous results we conclude that $\langle \delta_m \theta_r \rangle$ must also be IR finite at $\mathcal{O}(\beta)$. Arguments of this kind extend to all contributions to the galaxy power spectrum. 

At the 1-loop level, the above structure can be explicitly verified by calculating $\langle \delta_m \delta_r \rangle$ and $\langle \delta_m \theta_r \rangle$ with the help of the perturbative solutions up to third order presented in this work: we find that non-trivial cancellations between different diagrams ensure IR finiteness at $\mathcal{O}(\beta)$.

The same conclusions can be reached more directly in LPT. In this formalism, the cross power spectrum of two generic fluids $X$ and $Y$ is expressed as
\begin{equation}
P_{XY}(k) + (2\pi)^3 \delta^{(3)}(\vec{k}) \\ = \int \mathrm{d}^3 q\, e^{i \vec{q}\cdot \vec{k}} \exp \Big[-\frac{1}{2}k_i k_j A_{ij}^{XY}(\vec{q}) - \frac{i}{6} k_i k_j k_\ell W_{ij\ell}^{XY} (\vec{q})\Big] \,,
\end{equation}
where the cumulant expansion has been truncated at third order, as sufficient for a 1-loop calculation:
\begin{equation}
A_{ij}^{XY}(\vec{q}) = \langle \Delta_i \Delta_j \rangle_c \,,\qquad W_{ij\ell}^{XY} (\vec{q}) = \langle \Delta_i \Delta_j \Delta_\ell \rangle_c\,,
\end{equation}
with $\vec{\Delta} \equiv \bm{\psi}_Y (\vec{q}) - \bm{\psi}_X (0)\,$. The cumulant subscript $c$ indicates that only connected terms should be included. Focusing for example on the second cumulant, it can be written as
\begin{equation}\label{eq:second_cumulant}
A_{ij}^{XY}(\vec{q}) = \langle \psi^i_Y \psi^j_Y + \psi^i_X \psi^j_X - 2 \psi_X^i \psi_Y^j \rangle + 2\, \langle  \psi_X^i \psi_Y^j  -  \psi_X^i (0) \psi_Y^j (\vec{q})  \rangle\,,
\end{equation}
where the arguments were omitted for correlators of displacements evaluated at the same point. In the single-fluid case $X = Y$, the right-hand side of~\eqref{eq:second_cumulant} trivially vanishes for $\vec{q}\to 0$, manifesting the automatic absence of IR divergences characteristic of LPT. By contrast, for $X\neq Y$ the first term on the right-hand side does not vanish in the soft limit~\cite{Chen:2019cfu}; it is proportional to $|\bm{\psi}_X - \bm{\psi}_Y|^2$. The case of interest here, recalling~\eqref{eq:P_chib}, is $X = \chi$ and $Y = b$. We find that in the soft limit $P_{\chi b} (k)$ is proportional to $\psi_r^2 \sim \mathcal{O}(\beta^2)$. This implies that IR divergences are absent from $P_{\chi b}$, in addition to $P_\chi$ and $P_b\,$, at $\mathcal{O}(\beta)$. Analogous considerations apply to all other contributions to the LPT galaxy power spectrum.

In conclusion, the violation of the consistency relations in the galaxy power spectrum is postponed to $\mathcal{O}(\beta^2)$, where it carries negligible phenomenological impact.

\section{Statistical methods}\label{sec:methods}

\subsection*{Fisher matrix}

The Fisher matrix is a powerful tool to forecast how strongly future experiments can constrain the parameters of a given model, without having access to any kind of data (real or simulated). The idea is based on the properties of the Maximum Likelihood Estimator (MLE) of the parameters that we want to constrain \cite{Tegmark:1997rp}. We do not aim here for rigor, but to expose the intuition behind the Fisher matrix, and refer the reader not familiar with this formalism to well-known textbooks, \emph{e.g.} Ref.~\cite{Dodelson:2003ft}.

Consider a cosmological model described by a set of unknown parameters, collectively denoted as $\theta$, that we want to constrain. Let us assume that the real world is well described by the chosen model with a specific set of ``true'' parameters, denoted as $\theta_0$. Then an experiment will be able to measure a set of observables $\oexp_k$, labeled by the index $k$, for which the covariance matrix $C_{k k^\prime}$ is known. These experimental observables also have a known theoretical expression as a function of the parameters, $\oth_{k}( \theta )$. Assuming that the experiment we want to forecast has a Gaussian likelihood, its log-likelihood, given a realization of the data, is
\begin{equation}\label{eq:gauss_lik}
\mathcal{\chi}^2(\theta) = \left(\oexp_k-\oth_k (\theta)\right) \left(C^{-1}\right)_{k k^\prime}\left(\oexp_{k^\prime}-\oth_{k^\prime} (\theta)\right) \,,
\end{equation}
where sums over repeated indices are understood. The quantity in~\eqref{eq:gauss_lik} is maximized by the MLE $\hat{\theta}$, which asymptotically converges to the true $\theta_0$. Then we can expand the likelihood around this point, obtaining the following expression for the likelihood function
\begin{equation}\label{eq:expanded_lik}
\begin{split}
\chi^2 (\theta)  \simeq \chi^2 (\theta_0) 
& + (\theta-\theta_0)^i(\theta-\theta_0)^j \Big[ - \frac{\partial^2 \oth_k (\theta_0)}{\partial\theta_i \partial\theta_j} \left(C^{-1}\right)_{k k^\prime}\left(\oexp_{k^\prime}-\oth_{k^\prime} (\theta_0)\right)  + \frac{\partial\oth_k (\theta_0)}{\partial\theta_i}\left(C^{-1}\right)_{k k^\prime} \frac{\partial\oth_{k^\prime}(\theta_0)}{\partial\theta_j}  \Big]\,,
\end{split}
\end{equation}
where the first order term has been set to zero, since the likelihood derivative vanishes when evaluated at the MLE by definition, and we have used the fact that the only quantities that depend on the parameters are the $\oth$. If the sample is large enough, we can expect the likelihood to reach its expectation value. The only quantities in~\eqref{eq:expanded_lik} that are stochastic are the $\oexp$, and since we assume the theory to correctly model the data, 
\begin{equation}
\mathbb{E}\left[ \oexp\right] = \oth (\theta_0) \,.
\end{equation}
This implies that 
\begin{equation}
\mathbb{E}\left[\chi^2 (\theta)\right] = \chi^2 (\theta_0) + (\theta-\theta_0)^i(\theta-\theta_0)^j \frac{\partial\oth_k (\theta_0)}{\partial\theta_i} \left(C^{-1}\right)_{k k^\prime} \frac{\partial\oth_{k^\prime}(\theta_0)}{\partial\theta_j}  \,.
\end{equation}
The quantity 
\begin{equation}\label{eq:fisher_def}
\mathcal{F}_{ij} \equiv \frac{\partial\oth_k (\theta_0) }{\partial\theta_i}\left(C^{-1}\right)_{k k^\prime} \frac{\partial\oth_{k^\prime} (\theta_0) }{\partial\theta_j}
\end{equation}
is the Fisher matrix. Its importance lies in the fact that it is an estimator of the inverse covariance of the MLE $\hat{\theta}$, and hence it gives an estimate of the constraining power of a future survey. Since it does not depend on any data, it can be computed theoretically in a given model.

\subsection*{Estimators for power spectrum and bispectrum}

Here we derive simple analytical expressions for the signal-to-noise ratio of the relevant observables considered in this work, namely the galaxy power spectrum and bispectrum. We define the momentum space correlators as
\begin{equation*}
\langle\delta_g(\vec{k}_1)\ldots\delta_g(\vec{k}_n)\rangle\equiv(2\pi)^3\delta^{(3)}\bigg(\sum_{i\, = \, 1}^n\vec{k}_i\bigg)C_g^{(n)}(\vec{k}_1,\ldots,\vec{k}_n)\,,
\end{equation*}
where, in particular, $C^{(2)}_g(\vec{k}_1,\vec{k}_2)=P_g(k)$ is the power spectrum evaluated at $k=|\vec{k}_1|=|\vec{k}_2|$ and $C_g^{(3)}(\vec{k}_1,\vec{k}_2,\vec{k}_3)=B_g(k_1,k_2,k_3)$ is the bispectrum. Experimentally, correlation functions are measured by selecting each momentum $\vec{k}_i$ entering the correlation function inside a shell of radius $|\vec{k}_i|\equiv k_i$ and width $\delta k_i$. Given an $n$-point correlation function, a suitable estimator of the corresponding experimentally measured quantity is
\begin{equation}\label{eq:Bhat_n}
\hat{C}_g^{(n)}(\vec{k}_1,\ldots,\vec{k}_n)=\frac{1}{V\hat{V}_{1\ldots n}}\int_{k_{1\ldots n}}\mathrm{d}p_{1\ldots n}\,\delta_g(\vec{p}_1)\ldots\delta_g(\vec{p}_n)\,,
\end{equation}
where the integral over each momentum must be performed within the corresponding shell. $V$ is the volume of the survey, while the number of configurations in momentum space, $\hat{V}_{1\ldots n}$, is defined as
\begin{equation}
    \hat{V}_{1\cdots n} = \int_{k_{1\ldots n}}\mathrm{d}p_{1\ldots n} \equiv \int_{k_1 \pm \frac{\delta k_1}{2}} \frac{\mathrm{d}^3 p_1}{(2\pi)^3}\; \ldots\; \int_{k_n \pm \frac{\delta k_n}{2}} \frac{\mathrm{d}^3 p_n}{(2\pi)^3} \,(2\pi)^3 \delta^{(3)}\bigg(\sum_{i\, =\, 1}^n \vec{p}_i \bigg)\, .
\end{equation}
The values of $\hat{V}_{1\ldots n}$ needed here are $\hat{V}_{12}=k^2\delta k/(2\pi^2)$ and $\hat{V}_{123} = k_1 k_2 k_3\delta k_1\delta k_2\delta k_3/(8\pi^4)$. It is easy to show that the definition in~\eqref{eq:Bhat_n} is such that $\langle \hat{C}_g^{(n)}\rangle=C_g^{(n)}$. When comparing the effects of new physics on an $n$-point function with respect to $\Lambda$CDM, we also need an estimator of the variance in the $\Lambda$CDM correlation function. This is given by
\begin{equation}
\text{Var}\big(\hat{C}_g^{(n)}\big)=\frac{1}{V^2\hat{V}_{1\ldots n}^2}\int_{k_{1\ldots n}}\mathrm{d}p_{1\ldots n}\,\delta_g(\vec{p}_1)\ldots\delta_g(\vec{p}_n)\int_{k_{1\ldots n}}\mathrm{d}q_{1\ldots n}\,\delta_g(\vec{q}_1)\ldots\delta_g(\vec{q}_n)-(\hat{C}_g^{(n)})^2\,.
\end{equation}
Focusing only on the sensitivity to the dark force strength $\beta$ and recalling~\eqref{eq:fisher_def}, we thus arrive at the following expression for the one-dimensional Fisher matrix
\begin{equation}\label{eq:1DF}
\mathcal{F}_{\beta\beta} = \sum_{k_1 \geq\, \ldots\, \geq k_n}\frac{\big(\partial \hat{C}_g^{(n)}/\partial \beta\big)^2}{\text{Var}\big(\hat{C}_g^{(n)}\big)}\;,
\end{equation}
where the sum must be performed imposing $k_1\geq \ldots\geq k_n$ while satisfying the kinematic condition $\sum_{i\, =\, 1}^n\vec{k}_i = 0$. Using these definitions, we can provide simple semi-analytical estimates of the forecasted bounds on $\beta$ from the power spectrum and bispectrum. For simplicity we work in real space and consider only $b_1$ in the bias expansion, though the method can be straightforwardly generalized.

Making use of the results of Appendix~\ref{sec:eulerian} we write the galaxy power spectrum in real space, including 1-loop corrections, as
\begin{equation}\label{eq:Pg}
    P_g \simeq b_1^2\left(1+ 2 \varepsilon \fc \log\frac{a}{a_{\rm eq}}\right)P_{m,L}^{\rm CDM} + b_1^2\left(1+ 4\varepsilon \fc \log\frac{a}{a_{\rm eq}}\right) P_{m,{\rm 1\mbox{-}loop}}^{\rm CDM}\,,
\end{equation}
where only the leading log-enhanced corrections were included and the standard form of $P_{m,{\rm 1\mbox{-}loop}}^{\rm CDM}$ can be found~{\it e.g.} in Eqs.~(2) and (3) of Ref.~\cite{Senatore:2014via}. We have also neglected higher-order bias terms. The variance in $\Lambda$CDM is
\begin{equation}
\text{Var}\big(\hat{P}_g\big)=\frac{2}{V\hat{V}_{12}}\left[b_1^2\Big(P_{m,L}^{\rm CDM} + P_{m,{\rm 1\mbox{-}loop}}^{\rm CDM}\Big) + \frac{1}{\bar{n}_g}\right]^2\,,
\end{equation}
where the shot noise has been included as well. Hence the Fisher matrix has the expression
\begin{align}\label{eq:fisher_ps}
    \mathcal{F}_{\beta\beta}&=\left( 2 b_1^2 \mt^2\fc^2\log\frac{a}{a_{\rm eq}}\right)^2 V \sum_k \frac{\Big(P_{m,L}^{\rm CDM} + 2 P_{m,{\rm 1\mbox{-}loop}}^{\rm CDM}\Big)^2  \hat{V}_{12}}{2\left[b_1^2\Big( P_{m,L}^{\rm CDM} + P_{m,{\rm 1\mbox{-}loop}}^{\rm CDM}\Big)+\frac{1}{\bar{n}_g}\right]^2}\nonumber\\
    &\simeq\left( 2 b_1^2 \mt^2\fc^2\log\frac{1+z_{\rm eq}}{1+z}\right)^2V\int_{k_{\min}}^{k_{\max}}\frac{\mathrm{d}kk^2}{4\pi^2} \left(\frac{P_{m,L}^{\rm CDM}(k,z) + 2 P_{m,{\rm 1\mbox{-}loop}}^{\rm CDM}(k,z)}{b_1^2\Big(P_{m,L}^{\rm CDM}(k,z) + P_{m,{\rm 1\mbox{-}loop}}^{\rm CDM}(k,z)\Big)+\frac{1}{\bar{n}_g}}\right)^2\nonumber\\
    &= 2.1\times 10^6 \left(\mt^2\fc^2\log\frac{1+z_{\rm eq}}{1+z}\right)^2\,,
\end{align}
where Euclid forecast parameters for the $z=1.0$ bin~\cite{Euclid:2019clj} have been used for $b_1$, $\Bar{n}_g$ and $V$. We took $k_{\rm min} = 2\pi/V^{1/3}$ and $k_{\rm max} = k_{\rm NL,R}(z = 1) = 0.27\, h\, \mathrm{Mpc}^{-1}$, where $k_{\rm NL,R}$ is related to the rms value of the velocity field,
\begin{equation}
\langle v_m^i (\vec{x}, z) v_m^j (\vec{x}, z) \rangle \simeq \frac{\mathcal{H}^2 }{k_{\rm NL,R}(z)^2}\,\delta^{ij} \,,\qquad k_{\rm NL,R}(z) = \left( \int \frac{\mathrm{d}k}{6\pi^2} P_{m,L}^{\rm CDM} (k,z) \right)^{-1/2}\,.
\end{equation}
Since in the single-parameter case the forecasted $1\sigma$ uncertainty is simply $1/\sqrt{\mathcal{F}_{\beta\beta}}$, we arrive at the expected one-dimensional bound on $\beta$ from the Euclid power spectrum,
\begin{equation}\label{eq:1D_P}
    \beta \mt^2 \fc^2  < 1.8 \times 10^{-4}\; \qquad (\mathrm{power}\;\mathrm{spectrum},\;95\%\;\mathrm{c.l.}).
\end{equation}
An alternative approach consists in determining $k_{\rm max}$ as the scale where the size of the new physics correction $\Delta P_g$ in~\eqref{eq:Pg} equals the perturbative uncertainty on the 1-loop theoretical prediction in $\Lambda \mathrm{CDM}$. To do so, we take the following estimate of the theory error~\cite{Baldauf:2016sjb}
\begin{equation}
E_P(k, z) = b_1^2 \left( \frac{D_{1m}^{\rm CDM}(z)}{D_{1m}^{\rm CDM}(0)} \right)^4 P_{m,L}^{\rm CDM} (k, z) \left( \frac{k}{0.61\,h\,\mathrm{Mpc}^{-1}} \right)^3\,,
\end{equation}
where the parametric dependences on redshift and scale are those expected from the 2-loop terms, whereas the normalization is fixed to $E_P/(b_1^2 P_L) = 1\%$ at $k=0.2\,h\,\mathrm{Mpc}^{-1}$ and $z = 0.6$ based on the results of Ref.~\cite{Nishimichi:2020tvu}. We solve for $k_{\rm max}$ the equation $\Delta P_g = E_P$ (where, on the left-hand side, $\beta \mt^2 f_\chi^2 $ is evaluated at the $95\%$~c.l.~bound as a function of $k_{\rm max}$), obtaining for $z = 1$ that $k_{\rm max} = 0.20\, h\, \mathrm{Mpc}^{-1}$, corresponding to $\beta \mt^2\fc^2  < 2.5 \times 10^{-4}$, about $40\%$ weaker than~\eqref{eq:1D_P}. 

Next, we turn to the tree-level bispectrum. The dominant contribution comes from the $\Lambda$CDM-like piece in~\eqref{eq:bispectrum}, which yields
\begin{equation}
   \frac{\partial }{\partial\beta}B_g(k_1,k_2,k_3, a) \simeq 4 \,\mt^2\fc^2\log\frac{a}{a_{\rm eq}} B^{\rm CDM}_g (k_1,k_2,k_3, a),
\end{equation}
while the variance for scalene triangles is
\begin{equation}\label{eq:B_variance}
\text{Var}\big(\hat{B}_g \big) (k_1, k_2, k_3, z) = \frac{1}{V\hat{V}_{123}}\left(b_1^2P_{m,L}^{\rm CDM}(k_1,z)+\frac{1}{\bar{n}_g}\right)\left(b_1^2P_{m,L}^{\rm CDM}(k_2,z)+\frac{1}{\bar{n}_g}\right)\left(b_1^2P_{m,L}^{\rm CDM}(k_3,z)+\frac{1}{\bar{n}_g}\right).
\end{equation}
Defining $s_{123} = 1, 2, 6$ for scalene, isosceles, and equilateral triangles respectively, the Fisher matrix then reads 
\begin{align}\label{eq:fisher_bis}
    \mathcal{F}_{\beta\beta}&=\left(4\,\mt^2\fc^2\log\frac{a}{a_{\rm eq}}\right)^2\sum_{k_1\geq k_2\geq k_3}\frac{B^{\rm CDM}_g (k_1,k_2,k_3,a)^2}{s_{123}\text{Var}\big(\hat{B}_g\big) (k_1, k_2, k_3, a)}\nonumber \\
    &\simeq\left(4\,\mt^2\fc^2\log\frac{1 + z_{\rm eq}}{1 + z}\right)^2 \Bigg[\int_{\kmin}^{\kmax}\frac{\mathrm{d}k_1}{\kmin}\int_{k_1^*}^{k_1}\frac{\mathrm{d}k_2}{\kmin}\int_{k_2^*}^{k_2}\frac{\mathrm{d}k_3}{\kmin}\frac{B_g^{\rm CDM}(k_1,k_2,k_3,z)^2}{\text{Var}(\hat{B}_g)(k_1, k_2, k_3, z)} \nonumber \\
    &+\int_{\kmin}^{\kmax}\frac{\mathrm{d}k_1}{\kmin}\int_{\kmin}^{\kmax^*}\frac{\mathrm{d}k_2}{\kmin}\frac{B_g^{\rm CDM}(k_1,k_1,k_2,z)^2}{2\text{Var}\big(\hat{B}_g\big) (k_1, k_1, k_2, z)} +\int_{\kmin}^{\kmax}\frac{\mathrm{d}k_1}{\kmin}\frac{B_g^{\rm CDM}(k_1,k_1,k_1,z)^2}{6\text{Var}\big(\hat{B}_g\big)(k_1,k_1,k_1,z)}\Bigg] \nonumber \\
    &= 9.0 \times 10^4 \left(\mt^2\fc^2\log\frac{1+z_{\rm eq}}{1+z}\right)^2, 
    \end{align}
where $k_1^*=\max(\kmin,k_1/2)$, $k_2^*=\max(\kmin,k_1-k_2)$ and $\kmax^*=\min(\kmax,2 k_1)$. The same Euclid parameters were assumed as for the power spectrum above, except for the smaller $k_{\rm max} = 0.11\, h\, \mathrm{Mpc}^{-1}$, given that here we are working at tree level. We thus obtain the forecasted bound from the Euclid bispectrum,
\begin{equation}\label{eq:1D_B}
    \beta \mt^2 \fc^2  < 8.7 \times 10^{-4} \left( \frac{0.11 \,h\,\mathrm{Mpc}^{-1}}{k_{\rm max}} \right)^{2.2}\; \qquad (\mathrm{bispectrum},\;95\%\;\mathrm{c.l.}),
\end{equation}
where the extrapolation to $k_{\rm max} > 0.11\,h\,\mathrm{Mpc}^{-1}$ obtained from the tree-level calculation is shown explicitly. Taking this extrapolation at face value, we would find that for $k_{\rm max} = 0.23\,h\,\mathrm{Mpc}^{-1}$ the sensitivity on $\beta \fc^2 \mt^2$ matches the one of the power spectrum in~\eqref{eq:1D_P}. We regard this as a strong motivation to extend to 1-loop order the calculation of the bispectrum in the presence of dark forces.

In analogy to the power spectrum, we can alternatively determine $k_{\rm max}$ by making use of the error estimate for the tree-level bispectrum~\cite{Baldauf:2016sjb},
\begin{equation}
E_B (k_1, k_2, k_3, z) = b_1^3  \left( \frac{D_{1m}^{\rm CDM}(z)}{D_{1m}^{\rm CDM} (0)} \right)^2 3 B_{m}^{\rm CDM} (k_1, k_2, k_3, z) \left( \frac{(k_1 + k_2 + k_3)/3}{0.31\,h \,\mathrm{Mpc}^{-1}} \right)^{1.5}\,,
\end{equation}
where the parametric dependences approximate -- admittedly in a very crude way -- the $1$-loop corrections. We solve for $k_{\rm max}$ the equation $\Delta B_g = E_B$, assuming $k_1 + k_2 + k_3 = 2 k_{\rm max}$, and obtain $k_{\rm max} \approx 0.075\, h\, \mathrm{Mpc}^{-1}$, with $\beta  \mt^2 \fc^2 < 2.4 \,\times 10^{-3}$. This conservative bound is weaker by a factor $3$ compared to the result for $k_{\rm max} = 0.11\, h\, \mathrm{Mpc}^{-1}$ in~\eqref{eq:1D_B}.

Finally, we discuss the $1/p$ enhancement in the bispectrum of two tracers, which takes the form
\begin{equation}
B_g^{AAB}(\vec{p}, \vec{p}_1, \vec{p}_2) \simeq \left(1 + 4 \, \varepsilon f_\chi \log \frac{a}{a_{\rm eq}}\right) B_{g}^{AAB,\,\mathrm{CDM}} + \varepsilon \Delta B_g^{\rm p} + \varepsilon \Delta B_g^{\rm np}\,, 
\end{equation}
where the pole contribution $\varepsilon \Delta B_g^{\rm p}$ in the squeezed limit $\vec{p}\to 0$ was given in~\eqref{eq:pole_B_real}, whereas the no-pole contribution $\varepsilon \Delta B_g^{\rm np}$ contains terms that are neither log-enhanced nor $1/p\,$-$\,$enhanced and can be neglected here. To obtain a sensitivity estimate we make some simplifying assumptions. Considering only $b_1$ and $b_r$ in the bias expansion, we assume $b_1^A = b_1^B = b_1$ but different relative bias, $b_r^A > b_r^B$. We also neglect the shot noise (which is not unreasonable given our $k_{\rm max} = 0.11\,h\,\mathrm{Mpc}^{-1}$), so that the bispectrum variance for scalene triangles is simply given by~\eqref{eq:B_variance} without the $1/\bar{n}_g$ terms. Then the one-dimensional Fisher matrix, obtained summing over the momenta $p_1 \geq p_2 \gg p$ and including only scalene triangles for simplicity, reads
\begin{align}\label{eq:F_betabeta_AAB}
\mathcal{F}_{\beta\beta} \simeq \widetilde{m}_s^4 f_\chi^2\, V \int_{k_{\rm min}}^{k_{\rm max}}\mathrm{d}p \int_{p^\ast}^{k_{\rm max}}\mathrm{d}p_2 \int_{p_2}^{p_2^\ast}\mathrm{d}p_1 \frac{p\, p_1 p_2}{8\pi^4} \frac{\left(8 f_\chi \log \frac{1+z_{\rm eq}}{1+z} B_g^{AAB,\,\mathrm{CDM}} + \Delta B_g^{\rm p}\right) \Delta B_g^{\rm p} }{b_1^6\, P_{m,L}^{\rm CDM}(p)P_{m,L}^{\rm CDM}(p_1)P_{m,L}^{\rm CDM}(p_2)}
\end{align}
where $p_2^\ast = \mathrm{min}(k_{\rm max}, p_2 + p)$. Here $p^\ast = \mathrm{min}(r p, k_{\rm max})$ restricts the integral to configurations with minimum squeezing ratio equal to $r$ (namely $p_2 \geq r p$). In the numerator we have retained only the terms that are sensitive, either linearly or quadratically, to the $1/p$ pole,
\begin{align}
B_g^{AAB,\,\mathrm{CDM}}(p,p_1,p_2) \simeq&\; \frac{b_1^3}{7} P_{m,L}^{\rm CDM}(p) \left[ P_{m,L}^{\rm CDM}(p_1) \left( 13 + 8\frac{(p_1 - p_2)^2}{p^2}\right) - 7  \,  \frac{\partial P_{m,L}^{\rm CDM}(p_1)}{\partial \log p_1} \frac{(p_1 - p_2)^2}{p^2} \right]\,, \nonumber \\
\Delta B_g^{\mathrm{p}}(p,p_1,p_2) \simeq&\; P_{m,L}^{\rm CDM}(p) P_{m,L}^{\rm CDM}(p_1) \frac{7}{6} b_1^3 \bigg(\frac{ b_r^A - b_r^B }{b_1}\bigg) \frac{p_1 (p_1 - p_2)}{p^2} \,, 
\end{align}
which have been expanded for small $(p_1-p_2)/p_1 \sim \mathcal{O}(p/p_1)$. Setting the minimum squeezing ratio to $r = 5$, we obtain for the Euclid $z = 1.0$ bin that
\begin{equation}
\mathcal{F}_{\beta\beta} = \widetilde{m}_s^4 f_\chi^2 \frac{b_r^A - b_r^B}{b_1} \bigg[ 4.9 \times 10^3\, f_\chi \log \frac{1+ z_{\rm eq}}{1+z} + 1.3\times 10^3\, \frac{b_r^A - b_r^B}{b_1} \bigg]\,,
\end{equation}
showing that for $f_\chi \simeq 1$ the piece of~\eqref{eq:F_betabeta_AAB} proportional to $B_g^{AAB,\,\mathrm{CDM}} \Delta B_g^{\mathrm{p}}$ dominates the sensitivity. This corresponds to a forecasted constraint
\begin{equation}\label{eq:1D_P_multi}
    \beta \mt^2 \fc^{3/2}  < \frac{1.0 \times 10^{-2}}{\left( \frac{b_r^A - b_r^B}{b_1} \right)^{1/2}}\; \qquad (\mathrm{pole}\;\mathrm{of}\;\mathrm{multi}\text{-}\mathrm{tracer}\;\mathrm{bispectrum},\;95\%\;\mathrm{c.l.}).
\end{equation}
We thus find that, even assuming two tracers that differ enough to satisfy $(b_r^A - b_r^B)/b_1 \sim \mathcal{O}(1)$, this bound originating from the $1/p$ enhancement is one order of magnitude weaker than the constraint from the single tracer bispectrum,~\eqref{eq:1D_B}, which relies entirely on the enhanced growth of structure.

\section{Full parameter constraints}\label{sec:full_results}
\begin{figure}
    \centering 
   \includegraphics[scale = 0.7]{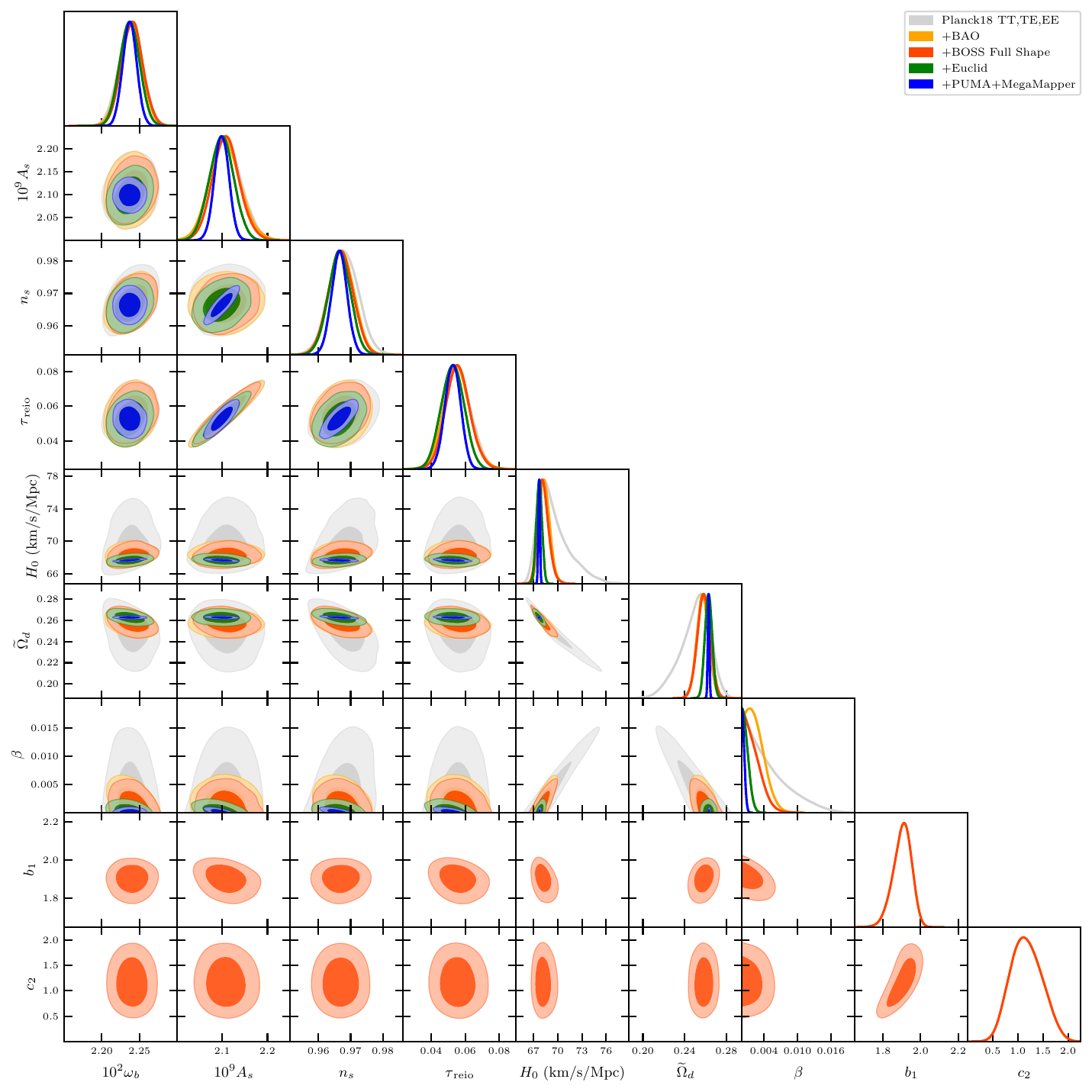}
    \caption{Full cosmological parameter constraints from the combination of Planck data with current and upcoming LSS data. CMB bounds alone are shown in gray, the inclusion of current BAO measurements in orange, and the further addition of the full shape of the BOSS galaxy power spectrum in red. The last combination provides the strongest bound on $\beta$ from the current data sets used in this work. To avoid clutter, we only plot the linear bias parameter $b_1$ and the quadratic bias parameter $c_2$ for the high-$z$ NGC BOSS sample. We also show the gain in constraining power that will come from future surveys like Euclid, in green, and PUMA and MegaMapper, in blue.}
    \label{fig:BOSS_full}
\end{figure}

In Fig.~\ref{fig:BOSS_full} we show the 9$\times$9 two-dimensional posterior distributions of the parameters in our model, extending Fig.~\ref{fig:money}. For simplicity, among the bias parameters we only show the linear bias $b_1$ and the quadratic bias $c_2$ (given by $c_2 = (b_1 + b_2/2 + 2 b_{K^2}/3)/\sqrt{2}$, see footnote~\ref{foot:bias}) of the high redshift bin of the north galactic cap (NGC) in BOSS. Compared to the Planck bounds, shown as gray contours, the LSS data mainly constrain the three-dimensional sub-space of $(\beta, H_0, \widetilde{\Omega}_d)$. Notice how the nonlinear bias is highly degenerate with $b_1$, whose uncertainty then becomes the limiting factor in the final constraint on $\beta$. 

\begin{figure}
    \centering   
    \includegraphics[scale = 0.7]{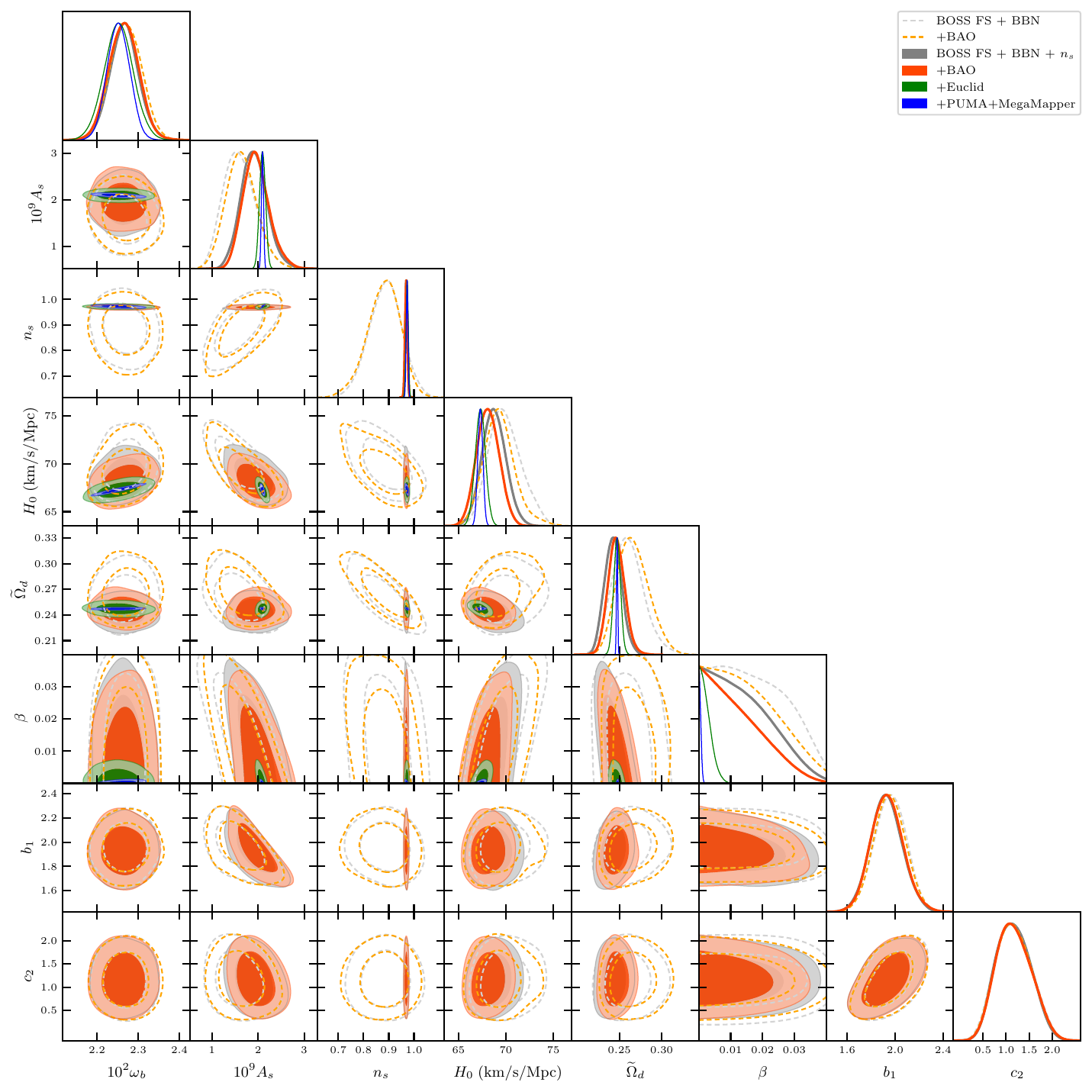}
    \caption{LSS-only constraints for the same parameters of Fig.~\ref{fig:BOSS_full} (except for $\tau_\mathrm{reio}$, which is fixed to $0.0549$).}
    \label{fig:lss_only_plot}
\end{figure}

We have also repeated the analysis presented in the main text using LSS-only likelihoods, \emph{i.e.}~removing the Planck bounds. This quantifies how strongly the model can be constrained with late-time data alone. We have added to all LSS likelihoods a Gaussian prior on $\omega_b$ from Big Bang Nucleosynthesis (BBN)~\cite{Aver:2015iza,Cooke:2017cwo},\begin{equation}
\omega_b = 0.02268 \pm 0.00036
\end{equation}
at $68\%$~c.l., and fixed $\tau_{\rm reio}$ to its Planck value (the latter is anyway irrelevant for the LSS bounds). The constraints are shown in Fig.~\ref{fig:lss_only_plot} for the BOSS+BBN~(+BAO) likelihoods with dashed light gray~(dashed orange) lines, and are dominated by the degeneracy between the various parameters, which are all quite uncertain. Notice that the prior on $\beta$ is limited at large values by the requirement that $\bar{s}> - 1/2$ at all times, thus avoiding a blow-up of the cosmological background solution caused by $\widetilde{m}_s = (1 + 2 \bar{s})^{-1}$ reaching a singularity. For the assumed negligible initial displacement, $\bar{s}_{\rm ini} \simeq 0$, the solution is well behaved provided $\varepsilon \lesssim 0.037$.

The addition of a Gaussian prior on $n_s$ from Planck, $n_s = 0.9698 \pm 0.0045$, dramatically improves the constraints on the standard cosmological model parameters. However, the bound on $\beta$ is still several times worse than the Planck one, see the dark gray and red sets of curves in Fig.~\ref{fig:lss_only_plot}. These results were expected, given the limiting constraining power of the BOSS data.

On the other hand, the upcoming measurements from Euclid will provide a competitive bound on $\beta$, independently of CMB data, as shown by the green contours in Fig.~\ref{fig:lss_only_plot}. Clearly, this will also be the case for the proposed Stage-IV surveys like PUMA and MegaMapper, shown in blue.

\begin{figure}
\centering
\includegraphics[scale = 0.48]{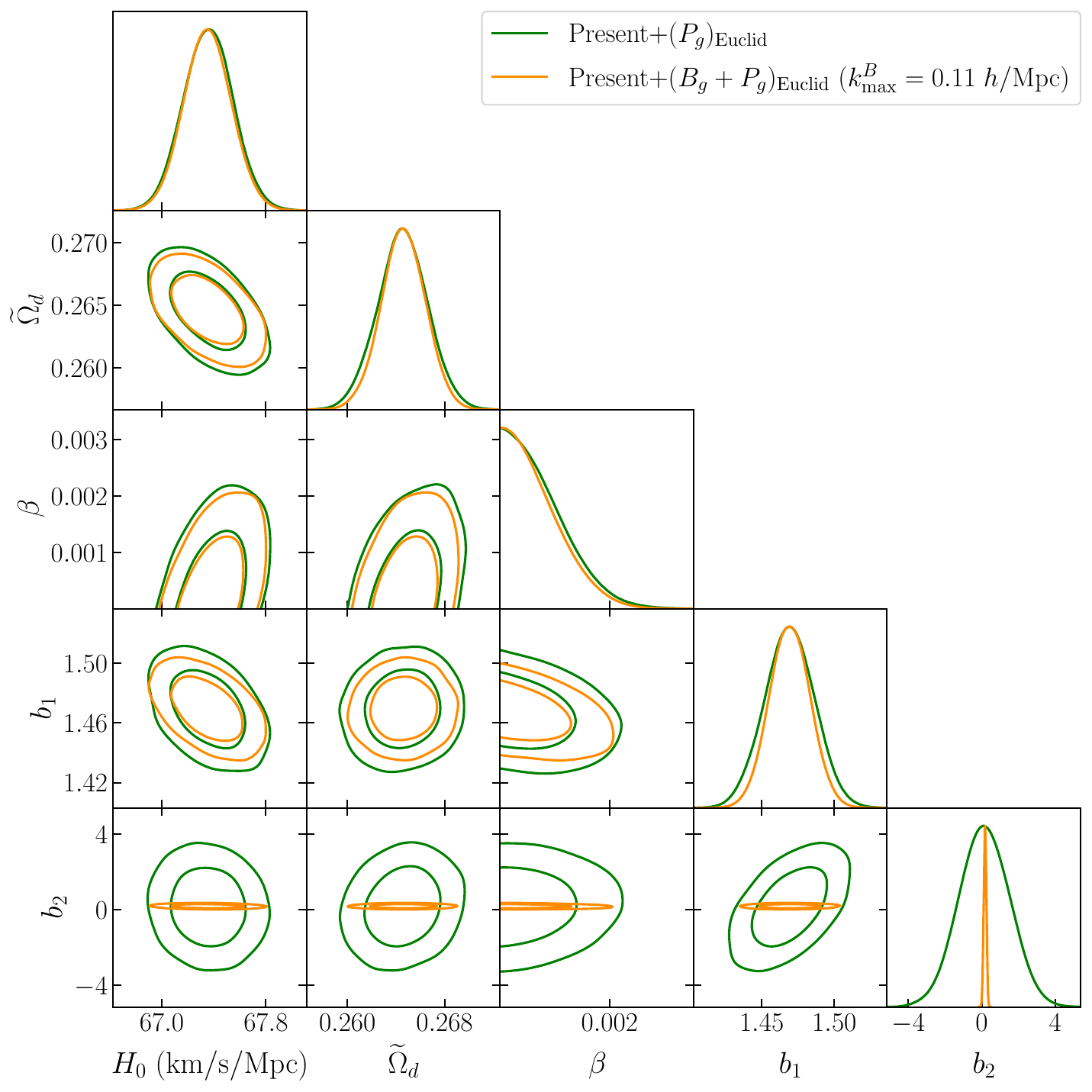}
\caption{The forecasted constraints from a joint analysis of the real space power spectrum and real space bispectrum for the Euclid survey, in combination with the current data (Planck+BAO+BOSS Full Shape). The green contours show the bounds from a power spectrum-only analysis, which is then supplemented with the measurement of the galaxy bispectrum up to $k_{\rm max} = 0.11\,h\,\mathrm{Mpc}^{-1}$, shown in orange. Here $b_1$ and $b_2$ refer to the linear and quadratic bias parameters of the lowest redshift bin in the Euclid spectroscopic sample.}\label{fig:euclid_bispectrum}
\end{figure}

In the main text, we discussed how the tree-level bispectrum does not add significant constraining power on $\beta$ to a CMB plus power spectrum analysis. This is further illustrated by Fig.~\ref{fig:euclid_bispectrum}. We show in green the bounds from the current data (Planck+BAO+BOSS Full Shape) plus our forecast for the real space Euclid galaxy power spectrum, to which we add, in orange, the real space Euclid bispectrum. In the figure, $b_1$ and $b_2$ refer to the linear and quadratic bias parameters for the bin with lowest redshift in the Euclid spectroscopic sample~\cite{Euclid:2019clj}. The bispectrum helps dramatically in better constraining nonlinear biases. However, the limited reach of the tree-level calculation used in our forecast -- we only consider triangular configurations up to $k_{\rm max} = 0.11\,h\,\mathrm{Mpc}^{-1}$ -- does not afford a significant improvement in the bound on $\beta$. 

\end{document}